\documentclass[conference,compsoc]{IEEEtran}
\IEEEoverridecommandlockouts
\def\isarxiv{1}
\PassOptionsToPackage{hyphens}{url}
\usepackage{cite}
\usepackage{amsmath,amssymb,amsfonts}
\usepackage{pifont}
\usepackage{algorithmic}
\usepackage{textcomp}
\usepackage[dvipsnames]{xcolor}
\usepackage{paralist}
\usepackage{enumitem}
\usepackage{hyperref}
\usepackage{cleveref}
\usepackage{float}
\usepackage[textsize=footnotesize]{todonotes}
\usepackage{glossaries}
\glsdisablehyper
\usepackage{caption}
\usepackage{subcaption}
\usepackage{listings}
\usepackage{xspace}
\usepackage{comment}
\usepackage{fancyvrb}
\usepackage{graphicx,txfonts}
\usepackage{multirow}
\usepackage{tikz}
\usepackage{multicol}
\usepackage{verbatim}
\usepackage{framed}
\usepackage{setspace}
\usepackage{cite}
\usepackage[tracking = true]{microtype}
\usepackage[T1]{fontenc}
\usepackage{bytefield}
\usepackage{booktabs}
\usepackage{xcolor}
\usepackage{mwe}
\usepackage{wrapfig}
\usepackage{color, colortbl}
\usepackage{threeparttable}
\usepackage{array}
\usepackage{adjustbox}
\usepackage{wasysym}
\usepackage{xparse}
\usepackage[htt]{hyphenat}

\input{listing-rust.sty}
\newif\ifabridged
\newif\ifnotabridged
\newif\ifanonymous
\newif\ifnotanonymous
\newif\ifarxiv
\newif\ifnotarxiv
\newif\ifreviewer
\newif\ifnotreviewer
\newif\iftags
\newif\ifnottags

\ifdefined\isabridged
\abridgedtrue
\fi

\ifabridged\notabridgedfalse
\else\notabridgedtrue
\fi

\ifdefined\isanonymous
\anonymoustrue
\fi

\ifanonymous\notanonymousfalse
\else\notanonymoustrue
\fi

\ifdefined\isarxiv
\arxivtrue
\fi

\ifarxiv\notarxivfalse
\else\notarxivtrue
\fi

\ifdefined\isreviewer
\reviewertrue
\fi

\ifreviewer\notreviewerfalse
\else\notreviewertrue
\fi

\ifdefined\hashtags
\tagstrue
\fi

\iftags\nottagsfalse
\else\nottagstrue
\fi

\newcommand{\colorbitbox}[3]{
  \sbox0{\bitbox{#2}{#3}}
  \makebox[0pt][l]{\textcolor{#1}{\rule[-\dp0]{\wd0}{\ht0}}}
  \bitbox{#2}{#3}
}

\makeatletter
\def\mathcolor#1#{\@mathcolor{#1}}
\def\@mathcolor#1#2#3{%
  \protect\leavevmode
  \begingroup
    \color#1{#2}#3%
  \endgroup
}
\makeatother

\newcommand{\newtext}[1]{{\color{red}{#1}}}
\newcommand{\newmath}[1]{\mathcolor{red}{#1}}

\iftags
\newcommand{\hashtagl}[1]{\marginpar{$\!\!\!\!\!\!\!\!\!\!$\color{blue}\em\##1\vspace{-7mm}}}
\newcommand{\hashtagr}[1]{\marginpar{$\!\!$\color{blue}\em\##1\vspace{-7mm}}}
\fi
\ifnottags
\newcommand{\hashtagl}[1]{}
\newcommand{\hashtagr}[1]{}
\fi

\newcommand{\dOne}{\ding{182}\xspace}
\newcommand{\dTwo}{\ding{183}\xspace}
\newcommand{\dThree}{\ding{184}\xspace}
\newcommand{\dFour}{\ding{185}\xspace}

\newcommand{\dCOne}{\ding{192}\xspace}
\newcommand{\dCTwo}{\ding{193}\xspace}
\newcommand{\dCThree}{\ding{194}\xspace}
\newcommand{\dCFour}{\ding{195}\xspace}
\newcommand{\dCFive}{\ding{196}\xspace}
\newcommand{\dCSix}{\ding{197}\xspace}
\newcommand{\dCSeven}{\ding{198}\xspace}
\newcommand{\dCEight}{\ding{199}\xspace}
\newcommand{\dCNine}{\ding{200}\xspace}
\newcommand{\dCTen}{\ding{201}\xspace}

\newcommand{\cmark}{\ding{51}}
\newcommand{\xmark}{\ding{55}}
\newlength{\dingwidth}
\newcommand{\dWdth}{\hspace{\dingwidth-.2em}}

\newcommand*\circled[1]{\tikz[baseline=(char.base)]{
            \node[shape=circle,draw,inner sep=1pt] (char) {\footnotesize#1};}}

\definecolor{mGreen}{rgb}{0,0.6,0}
\definecolor{mGray}{rgb}{0.5,0.5,0.5}
\definecolor{lGray}{rgb}{0.9,0.9,0.9}
\definecolor{mPurple}{rgb}{0.58,0,0.82}
\definecolor{backgroundColour}{rgb}{0.95,0.95,0.92}

\newcommand\LSTSize{\fontsize{7}{7.2}\selectfont}
\newcommand*\LSTfont{\LSTSize\ttfamily\SetTracking{encoding=*}{-0}\lsstyle}

\newcommand\realnumberstyle[1]{#1}

\makeatletter
\newcommand{\zebra}[2]{%
    {\realnumberstyle{#2}}%
    \begingroup
    \lst@basicstyle
    \ifodd\value{lstnumber}%
        \color{#1}%
        \rlap{\hspace*{\lst@numbersep}%
        \color@block{\linewidth}{\ht\strutbox}{\dp\strutbox}%
        }%
    \fi
    \endgroup
}
\makeatother
\newcommand{\breakingperiod}{
\nobreak\hspace{0pt}.\penalty0
}

\ExplSyntaxOn{}

\NewDocumentCommand{\longword}{m}
 {
  \texttt
   {
    \seq_set_split:Nnn \l_michael_lw_seq { . } { #1 }
    \seq_use:Nn \l_michael_lw_seq { \breakingperiod }
   }
 }

\ExplSyntaxOff{}

\lstdefinestyle{CStyle}{
    backgroundcolor=\color{backgroundColour},
    commentstyle=\color{mGreen},
    keywordstyle=\color{magenta},
    numberstyle=\tiny\zebra{lGray},
    stringstyle=\color{mPurple},
    basicstyle=\LSTfont,
    breakatwhitespace=false,
    breaklines=true,
    captionpos=b,    
    escapeinside={\%*}{*)},
    keepspaces=true,
    numbers=left,
    numbersep=5pt,
    showspaces=false,
    showstringspaces=false,
    showtabs=false,
    tabsize=2,
    language=C,
    classoffset=1,
    morekeywords={enter_domain, __writebeforeread, sdrob_call,sdrob_enter,sdrob_malloc,sdrob_exit,sdrob_init,sdrob_destroy,sdrob_free,sdrob_deinit,sdrob_dprotect, sdrad_call,sdrad_enter,sdrad_malloc,sdrad_exit,sdrad_init,sdrad_destroy,sdrad_free,sdrad_deinit,sdrad_dprotect},
    keywordstyle=\color{orange},
    classoffset=2,
    morekeywords={EXECUTION_DOMAIN,ISOLATED_DOMAIN,RETURN_TO_CURRENT,DATA_DOMAIN,SUCCESSFUL_RETURNED,EXECUTION_DOMAIN,NONISOLATED,RETURN_HERE,RETURN_TO_PARENT,NO_HEAP_MERGE,HEAP_MERGE,OK,OK,MALLOC_FAILED,WRITE_ENABLE,READ_ENABLE,ACCESSIBLE,INACCESSIBLE,ACCESSIBLE_DOMAIN,INACCESSIBLE_DOMAIN,udi_t},
    keywordstyle=\color{brown},
    classoffset=0,
}
\crefname{lstlisting}{listing}{listings}
\Crefname{lstlisting}{Listing}{Listings}

\presetkeys{todonotes}{fancyline}{}
\Crefname{section}{\S$\!$}{\S\S$\!$}

\newcounter{tncnt}

\newcounter{mgcnt}

\newcounter{jtcnt}

\newcounter{hecnt}

\renewcommand{\paragraph}{\noindent\textbf}

\def\BibTeX{{\rm B\kern-.05em{\sc i\kern-.025em b}\kern-.08em
    T\kern-.1667em\lower.7ex\hbox{E}\kern-.125emX}}
\newcommand{\CC}{\glsentrydesc{cc}\xspace}
\newcommand{\CCs}{\glsentrylongpl{cc}\xspace}
\newcommand{\CPLong}{\glsentrylongpl{cp}\xspace}
\newcommand{\CP}{\gls{cp}\xspace}
\newcommand{\CPs}{\glspl{cp}\xspace}
\newcommand{\WriteBeforeRead}{\texttt{Write-before-Read}\xspace}
\newcommand{\WriteBeforeReadOnly}{\texttt{Write-before-Read-Only}\xspace}
\newcommand{\WriteBeforeExecute}{\texttt{Write-before-Execute}\xspace}
\newcommand{\WriteBeforeExecuteOnly}{\texttt{Write-before-Execute-Only}\xspace}
\newcommand{\WriteOnce}{\texttt{Write-Once}\xspace}
\newcommand{\ReadOnce}{\texttt{Read-Once}\xspace}
\newcommand{\ExecuteOnce}{\texttt{Execute-Once}\xspace}
\newcommand{\Read}{\texttt{R}\xspace}
\newcommand{\Write}{\texttt{W}\xspace}
\newcommand{\Execute}{\texttt{X}\xspace}

\newcommand{\CCSWPerms}{\ensuremath{p_{sw}}\xspace}
\newcommand{\CCHWPerms}{\ensuremath{p_{hw}}\xspace}
\newcommand{\CCPermissions}{\ensuremath{p}\xspace}
\newcommand{\CCFlags}{\ensuremath{f}\xspace}
\newcommand{\CCOpPerms}{\ensuremath{p_{op}}\xspace}
\newcommand{\CCOType}{\texttt{otype}\xspace}
\newcommand{\CCExponent}{\ensuremath{E}\xspace}
\newcommand{\CCInternalExponent}{\ensuremath{I_E}\xspace}
\newcommand{\CCAddress}{\ensuremath{a}\xspace}
\newcommand{\CCMantissaWidth}{\ensuremath{MW}\xspace}
\newcommand{\CCConditional}{\ensuremath{c}\xspace}

\newcommand{\CCTop}{\ensuremath{T}\xspace}
\newcommand{\CCAllBitsTop}{\ensuremath{\CCTop\left[13:0\right]}\xspace}
\newcommand{\CCBottomBitsTop}{\ensuremath{\CCTop\left[2:0\right]}\xspace}
\newcommand{\CCUpperBitsTop}{\ensuremath{\CCTop\left[13:12\right]}\xspace}
\newcommand{\CCExpZeroTop}{\ensuremath{\CCTop\left[11:0\right]}\xspace}
\newcommand{\CCExpNonZeroTop}{\ensuremath{\CCTop\left[11:3\right]}\xspace}
\newcommand{\CCExpHighPart}{\ensuremath{\CCTop_E}\xspace}

\newcommand{\CCBase}{\ensuremath{B}\xspace}
\newcommand{\CCAllBitsBase}{\ensuremath{\CCBase\left[13:0\right]}\xspace}
\newcommand{\CCBottomBitsBase}{\ensuremath{\CCBase\left[2:0\right]}\xspace}
\newcommand{\CCUpperBitsBase}{\ensuremath{\CCBase\left[13:12\right]}\xspace}
\newcommand{\CCExpZeroBaseBitsToCompare}{\ensuremath{\CCBase\left[11:0\right]}\xspace}

\newcommand{\CCExpNonZeroBase}{\ensuremath{\CCBase\left[13:3\right]}\xspace}
\newcommand{\CCExpLowPart}{\ensuremath{\CCBase_E}\xspace}
\newcommand{\moncheri}{Mon~CHÉRI\xspace}

\newcommand{\CCOp}{\ensuremath{O}\xspace}
\newcommand{\CCAllBitsOp}{\ensuremath{\CCOp\left[13:0\right]}\xspace}

\newcommand{\CCBottomBitsOp}{\ensuremath{\CCOp\left[2:0\right]}\xspace}

\newcommand{\CCExpNonZeroOp}{\ensuremath{\CCOp\left[13:3\right]}\xspace}
\newcommand{\CCExtOp}{\ensuremath{\CCOp_{E}}\xspace}
\newcommand{\CCExpZeroExtOpHighPart}{\ensuremath{\CCExtOp\left[4:2\right]}\xspace}

\newcommand{\LCarry}{\ensuremath{L_{carry\_out}}\xspace}
\newcommand{\LMsb}{\ensuremath{L_{msb}}\xspace}

\newcommand{\TrivialAutoVarInit}{\texttt{-ftrivial-auto-var-init}\xspace}
\newcommand{\CheriWriteBeforeRead}{\texttt{-cheri-write-before-read}\xspace}

\newcommand{\CheriFreeRtos}{CheriFreeRTOS\xspace}
\newcommand{\CheriFlute}{CHERI-Flute64\xspace}
\newcommand{\MonCheriFlute}{MonCHÉRI-Flute64\xspace}

\newcommand{\OverheadRelativeToCheri}{$\approx$~3.5\%\xspace}
\newcommand{\OverheadRelativeTobaseline}{7\%}
\newcommand{\OverheadRelativeTobaselineX}{\OverheadRelativeTobaseline\xspace}
\newacronym{address}{\CCAddress}{address}
\newacronym{alu}{ALU}{arithmetic logic unit}
\newacronym{api}{API}{application programming interface}
\newacronym{bsv}{BSV}{Bluespec System Verilog}
\newacronym{cve}{CVE}{Common Vulnerability Enumeration}
\newacronym{cwe}{CVE}{Common Weakness Enumeration}
\newacronym[longplural={conditional capabilities}]{cc}{CC}{conditional capability}
\newacronym{cp}{CP}{conditional permission}
\newacronym{aslr}{ASLR}{address space layout randomization}
\newacronym{base}{\ensuremath{b}}{base}
\newacronym{cheri}{CHERI}{Capability Hardware Enhanced RISC Instructions}
\newacronym{dpc}{DPC++}{Data Parallel C++}
\newacronym{ie}{\CCInternalExponent}{internal exponent}
\newacronym{ip}{IP}{intellectual property}
\newacronym{ir}{IR}{intermediate representation}
\newacronym{jit}{JIT}{just-in-time}
\newacronym{exponent}{\CCExponent}{exponent}
\newacronym{fpga}{FPGA}{field-programmable gate array}
\newacronym{fpu}{FPU}{floating point unit}
\newacronym{gep}{GEP}{\texttt{GetElementPtr}}
\newacronym{gm}{g.m.}{geometric mean}
\newacronym{isa}{ISA}{instruction-set architecture}
\newacronym{lam}{LAM}{Linear Address Masking}
\newacronym{lut}{LUT}{lookup table}
\newacronym{lowerbound}{LB}{lower bound}
\newacronym{mmu}{MMU}{memory management unit}
\newacronym{mpu}{MPU}{memory protection unit}
\newacronym{msrc}{MSRC}{Microsoft Security Response Center}
\newacronym{msvc}{MSVC}{Microsoft Visual C++}
\newacronym{mw}{\CCMantissaWidth}{\emph{mantissa width}}
\newacronym{nist}{NIST}{National Institute of Standards and Technology}
\newacronym{upperbound}{UB}{upper bound}
\newacronym{os}{OS}{operating system}
\newacronym{optop}{\ensuremath{o}}{operation top}
\newacronym{opbound}{OB}{operation bound}
\newacronym{sast}{SAST}{static application security testing}
\newacronym{sdlc}{SDLC}{secure software development lifecycle}
\newacronym{ssa}{SSA}{static single-assignment}
\newacronym{top}{\ensuremath{t}}{top}
\newacronym{tbi}{TBI}{Top Byte Ignore}
\newacronym{tlsf}{TLSF}{Two-Level Segregated Fit}
\newacronym{pcc}{PCC}{Program Counter Capability}
\newacronym{uai}{UAI}{Upper Address Ignore}
\newacronym{xom}{XOM}{eXecute-Only-Memory}

\makeatletter
\def\bstctlcite{\@ifnextchar[{\@bstctlcite}{\@bstctlcite[@auxout]}}
\def\@bstctlcite[#1]#2{\@bsphack
  \@for\@citeb:=#2\do{%
    \edef\@citeb{\expandafter\@firstofone\@citeb}%
    \if@filesw\immediate\write\csname #1\endcsname{\string\citation{\@citeb}}\fi}%
  \@esphack}
\makeatother

\begin{document}
\bstctlcite{IEEEtran:BSTcontrol}
\title{Mon CHÉRI: Mitigating Uninitialized Memory Access with Conditional Capabilities}
\ifnotanonymous
\author{\IEEEauthorblockN{Merve G\"{u}lmez\textsuperscript{*,$\dagger$}, Håkan Englund\textsuperscript{*}, Jan Tobias M\"uhlberg\textsuperscript{$\ddagger$}, Thomas Nyman\textsuperscript{§}
\IEEEauthorblockA{\textit{\textsuperscript{*}Ericsson Security Research, \textsuperscript{$\dagger$}DistriNet, KU Leuven, \textsuperscript{$\ddagger$}Universit\'e Libre de Bruxelles, \textsuperscript{§}Ericsson Product Security} \\
\{merve.gulmez,hakan.englund,thomas.nyman\}@ericsson.com, jan.tobias.muehlberg@ulb.be}}}

\fi
\maketitle
\begin{abstract}
Up to 10\% of memory-safety vulnerabilities in languages like C and C++ stem from uninitialized variables.
This work addresses the prevalence and lack of adequate software mitigations for uninitialized memory issues, proposing architectural protections in hardware.
Capability-based addressing, such as the University of Cambridge's \acrshort{cheri}, mitigates many memory defects, including spatial and temporal safety violations at an architectural level.
\acrshort{cheri}, however, does not handle undefined behavior from uninitialized variables.
We extend the \acrshort{cheri} capability model to include ``\emph{\CCs}'', enabling memory-access policies based on prior operations.
This allows enforcement of policies that satisfy memory-safety objectives such as ``\emph{no reads to memory without at least one prior write}'' (\WriteBeforeRead{}).
We present our architecture extension, compiler support, and detailed evaluation of our approach on the QEMU full-system simulator and a modified \acrshort{fpga}-based CHERI-RISCV softcore.
Our evaluation shows \CCs are practical, with high detection accuracy while adding a small (\OverheadRelativeToCheri) overhead which is comparable to the cost of baseline  \acrshort{cheri} capabilities.

\end{abstract}
\section{Introduction}\label{sec:introduction}
\emph{Uninitialized variables}, variables that are declared but not
assigned a value, are a well-known source of software defects in the
C-family of programming languages.
In general, all run-time allocated memory in C, C++, and even in modern
languages like Rust, starts out as \emph{uninitialized}. 
In this state, the value of the memory is indeterminate and may not reflect
a valid state for the variable type.
Attempting to interpret uninitialized memory results in
\emph{undefined behavior}, which can cause security vulnerabilities.
In particular, uninitialized memory may expose residual data
from previously deallocated data structures.
This may inadvertently result in information disclosure, ranging from leaked
pointer values~\cite{Cho20} to the exposure of cryptographic keys~\cite{Sullivan14a}.
Leaked pointer values can be exploited by attackers to bypass \emph{\gls{aslr}}~\cite{Cho20}, while the use 
of uninitialized variables can enable arbitrary code execution attacks~\cite{Lu17}.
Following recent surveys, use-before-initialized conditions account for a
sizable 10\% of memory-safety vulnerabilities in the
wild~\cite{Joly20,Bialek20,Sutter24}.

Hardware-assisted defenses against memory-safety issues~\cite{Zhao24} are
motivated by the need to reduce the overhead of run-time defenses through
hardware/software co-designs and willingness by processor manufacturers to
incorporate
mechanisms for software security into their designs~\cite{Qualcomm17, Arm19, Intel20}.
A prominent example is \emph{\gls{cheri}}~\cite{Woodruff14}, a joint research project of SRI International and the University of Cambridge.
\gls{cheri} extends
\glspl{isa} with
fine-grained memory protection and software compartmentalization.

In its default configuration, \gls{cheri} provides spatial safety through capability-based
addressing.
While this allows software to preclude many classes of
memory-safety defects, \gls{cheri}
does not address uninitialized variables.
\gls{msrc} conducted a comprehensive analysis of vulnerabilities reported in 2019 to assess the \gls{cheri} ISAv7~\cite{Joly20}. 
The findings indicate that only 31\% of the reported
vulnerabilities could have been mitigated through the default configuration
of \gls{cheri}.
An additional 24\% could be mitigated by \gls{cheri} when configured to provide
partial temporal safety under the Cornucopia
mechanism~\cite{WesleyFilardo20}. The
analysis further highlights that at least 12\% of the
assessed vulnerabilities could have been mitigated if \gls{cheri} would protect against
uninitialized access. This work explores and
evaluates extensions for \gls{cheri} to mitigate those 12\%.
\begin{figure*}[t]
    \begin{bytefield}[bitwidth=0.6em]{1}
        \begin{rightwordgroup}{\dCOne{} 1-bit Validity tag}
            \colorbitbox{black}{1}{}
        \end{rightwordgroup} \\
        \bitbox[]{1}{} \\
    \end{bytefield}
    \quad
    \begin{bytefield}[bitwidth=0.6em]{64}
        \bitbox{17}{\dCTwo{} Permissions} & \colorbitbox{lightgray}{2}{} &
        \bitbox{18}{\dCThree{} Object type} & \bitbox{27}{\dCFour{} Bounds} \\
        \bitbox{64}{\dCFive{} Baseline architecture address} \\
    \end{bytefield}
    \caption{In-memory representation of \gls{cheri} capabilities adapted from Watson et al.~\cite{Watson19}}\label{fig:chericap}
\end{figure*}

\paragraph{This paper and contributions.}
In this paper, we extend the \gls{cheri} capability
model to express memory-access policies that eliminate undefined behavior
associated with uninitialized memory.
We introduce \emph{\CPs} to capability-based
addressing. \CPs{} can express memory-access policies that take
previous operations on memory into account.
This enables \emph{\CCs} with fine-grained policies that satisfy
different instances of memory-safety objectives at different granularities, such
as \emph{``no reads of memory which has not been the subject of at least
one write''} (\WriteBeforeRead{}) or \emph{``this memory can be written to
only once''} (\WriteOnce{}).
We describe these \CPs{} in~\cref{sec:design}.
\CPs{} are enforced by introducing the notion of
\emph{operation-specific bounds} to capability-based addressing:

\begin{itemize}
    \item We introduce \emph{\CPLong} and \emph{\CCs} for capability-based addressing (\Cref{sec:design}).
    \item We integrate \CCs{} to CHERI-RISC-V in prototypes based on the QEMU full-system emulator and an FPGA softcore based on Flute64Cheri \acrshort{ip} (\Cref{sec:implementation}). 
    \item We add support for \WriteBeforeRead{} \CPLong{} to the \gls{cheri}-enabled Clang/LLVM compiler (\Cref{sec:instrumentation}) and embedded memory allocators (\Cref{sec:runtime}).
    \item We evaluate \CPs{} using $>1000$ \acrshort{nist} Juliet test suite test cases for uninitialized variables and using EEMBC CoreMark performance benchmarks (\Cref{sec:evaluation}).
\end{itemize}

Our results show that \CPs{} achieve 100\% detection with only six false positives ($\approx$~1\%) for the Juliet test suite, where the false positives do exhibit uninitialized (but non-vulnerable) accesses.
We further show that \CPs{} on CHERI-RISC-V impose only
\OverheadRelativeToCheri{} performance overhead in addition to that added
by \gls{cheri}, compared to benchmarks on an unmodified RISC-V softcore (\OverheadRelativeTobaselineX{} combined).
Consequently, \CP{} overhead is comparable to that of \gls{cheri}.
In summary, our work effectively and efficiently addresses uninitialized memory vulnerabilities, combining very high detection accuracy with performance penalties that are substantially lower than other detection mechanisms in hardware or software.
Our \CC-enhanced QEMU and toolchain protototypes, and evaluation artifacts are
available at \url{https://github.com/conditionalcapabilities}.

\section{Background}\label{sec:background}
\paragraph{How prevalent are uninitialized memory vulnerabilities?}
The prevalence of uninitialized memory as a source of vulnerabilities has been highlighted
in statistics based on the Common Vulnerability Enumeration (CVE) program.
\gls{msrc} reports that, between 2017 and 2019, uninitialized memory
vulnerabilities accounted for $\approx$~5--10\% of the \glspl{cve} issued
by Microsoft\cite{Joly20,Bialek20}. In 2019,
uninitialized memory accesses were the fourth largest class of memory
defects (10\%), after spatial (44\%), temporal (29\%), and type confusion
(14\%).  Similarly, a more recent survey of memory-safety CVEs between 2015
-- 2022 by Sutter~\cite{Sutter24} indicates that use-before-initialized
conditions account for $\approx$~9\% of all reported memory-safety
vulnerabilities, after lifetime safety (49\%), bounds safety (18\%), and 
type confusion (11\%), indicating that uninitialized memory accesses
remain an issue in industrial code bases that impacts system security.

\subsection{Capability-Based Addressing}\label{sec:capabilities}

\emph{Capability-based addressing} is a memory access-control paradigm originating from mainframe computers of the late 1950s and 60s~\cite{Levy84}.
In capability-based addressing, conventional references to locations in computer memory, i.e.\ \emph{pointers}, are replaced by protected objects called \emph{capabilities}~\cite{Dennis66}.
Capabilities carry, in addition to the referenced memory address,
additional permission information that is used by the processor or
memory-management subsystem to determine whether the accesses
performed through a capability are allowed.
While the exact composition of a capability can vary between different hardware instantiations, virtually all capability-based addressing schemes express allowed operations through at least \emph{Read} (\Read{}), \emph{Write} (\Write{}), and \emph{Execute} (\Execute{}) permissions that control whether references through a capability are permitted for load and store instructions or instruction fetches respectively.
Capabilities also include \emph{bounds information} that limit the range of memory that can be referenced via a particular capability.

Interest in capability-based addressing diminished with the introduction of \glspl{mmu} that, in addition to performing address translation between virtual and physical memory addresses, also manage access control to virtual memory.
However, capabilities differ fundamentally from the access control in \glspl{mmu}: whereas \glspl{mmu} associate permissions to \emph{individual memory pages}, capabilities associate permissions to the \emph{references used to address memory}.

\subsection{The \gls{cheri} Capability Architecture}\label{sec:cheri}

\gls{cheri}, which stands for \glsdesc{cheri}, is an \gls{isa} extension
for a capability-based architectural protection model and
hardware-software co-design. 
The \gls{cheri} architecture extends an underlying conventional \gls{isa} with hardware-supported capabilities that are used to protect virtual addresses used as code or data pointers.
The \gls{cheri} \gls{isa} specification~\cite{Watson23} defines the representation of capabilities held in registers and memory, as well as capability-aware instructions to manipulate them. 
Currently, implementations of CHERI exist for MIPS, RISC-V, and Armv8-A instruction sets.
\gls{cheri}-enabled processors have been developed by
Arm~\cite{Grisenthwaite22}, Microsoft~\cite{Amar23}, as well as in the RISC-V ecosystem.

\Cref{fig:chericap} shows the in-memory representation of a \gls{cheri} capability. 
Each capability is double the width of the native integer pointer type of the baseline architecture: 128~bits on 64-bit platforms and 64~bits on 32-bit platforms.
One additional bit, the \emph{validity tag} \dCOne, is stored separate from the capability and
is protects its integrity: any manipulation of the capability in-memory by non-capability-aware instructions invalidates its tag.
Capability-aware instructions maintain the tag as long as certain architectural invariants are met.
This prevents direct in-memory manipulations and injection of arbitrary data as capabilities.
The \emph{permissions} \dCTwo{} control how the capability can be used and consists of the permissions described in \Cref{sec:capabilities}.
The \emph{object type} \dCThree{} allows capabilities to be temporarily ``sealed'', which renders the capability unusable until it is ``unsealed'' by a special instruction.
Sealing is used by \gls{cheri} to implement opaque pointer types and fine-grained in-process isolation.
The \emph{bounds} \dCFour{} describe a lower and upper bound relative to the baseline architecture address \dCFive, which limits the portion of address space the capability is able to access.
To reduce the in-memory footprint of capabilities, the bounds are stored in a compressed format~\cite{Woodruff19} with both bounds in 28~bits (for a 64-bit address), loosing precision as the object size increases.

New capabilities in the \gls{cheri} architecture are always derived from an existing capability. The heritage of all capabilities can thus be traced back to the initial capabilities made available to firmware at boot time. 
\gls{cheri} enforces \emph{monotonicity} on newly created capabilities, ensuring that capabilities constructed by a capability-aware instruction cannot possess permissions or bounds that exceed those of the original capability.
The only exceptions to capability monotonicity are facilities for exception handling and compartmentalization using sealed capabilities, which allow non-monotonicity in a controlled manner to enable software to gain access to additional data capabilities.
 
The bounds information stored together with the virtual address forms the basis for the memory-safety properties provided by \gls{cheri}.
Each allocation made by a program running on a \gls{cheri}-capable processor is associated with a capability that describes, in addition to the address, the valid bounds of the object (or sub-object) in memory.
This allows \gls{cheri} to provide inherent spatial memory-safety properties.
Extensions to the \gls{cheri} software stack have explored adding temporal-safety properties to heap-based allocations~\cite{Xia19,WesleyFilardo20,Filardo24} and sandboxing~\cite{Chisnall17}.

\gls{cheri} does not enforce type safety for capabilities, nor does it prevent software from accessing uninitialized memory using a capability it possesses.
Consequently, \gls{cheri} must be complemented with compiler-based type-safety analysis, and instrumentation passes that zero local variables before first use~\cite{Bialek20,Milburn17} as well as heap allocators returning zeroed memory. 
Security analyses of CHERI (e.g., by \gls{msrc}\cite{Joly20})  generally take such mitigations for granted.

\section{Conditional Capability Design}\label{sec:design}
\glsreset{cp}

%As a motivating example,
Consider the uninitialized pointer-dereference vulnerability in the Linux kernel shown in \Cref{lst:example}.
This vulnerability allows control-flow hijacking due to the uninitialized \texttt{backlog} pointer \dOne{} being dereferenced at \dFour{} when \mbox{\texttt{cpg->eng\_st != ENGINE\_IDLE}}~\dTwo.
An attacker could exploit this vulnerability to achieve arbitrary code execution by spraying the stack to take control of the value of \texttt{backlog} and make it point to attacker-controlled code~\cite{Lu17}.

Our goal is to detect the first use of the uninitialized pointer in the \texttt{if}-clause at \dThree. Unlike methods that automatically zero out memory~\cite{Milburn17,Bialek20},
\CPs{} offer the same protection against uninitialized memory use but with smaller and more predictable run-time overhead.

\begin{lstlisting}[float=t, style=CStyle, label={lst:example}, caption={Excerpt from Linux' \texttt{queue\_manag()} function defined in \texttt{drivers/crypto/mv\_cesa.c} showing an uninitialized pointer dereference patched in April 2015~\cite{King15}.}]
static int queue_manag(void *data){
    cpg->eng_st = ENGINE_IDLE;
    do {
      %*\dWdth*)struct crypto_async_request *async_req = NULL;
      %*\dOne*)struct crypto_async_request *backlog;
      %*\dWdth*)/* ... cpg->eng_state may change state here ... */
      %*\dTwo*)if (cpg->eng_st == ENGINE_IDLE) {
      %*\dWdth*)   backlog = crypto_get_backlog(&cpg->queue);
      %*\dWdth*)   /* ... */
      %*\dWdth*)}

      %*\dThree*)if (backlog) {
      %*\dFour*)   backlog->complete(backlog, -EINPROGRESS);
      %*\dWdth*)}
      %*\dWdth*)/* ... */
    } while (!kthread_should_stop());
    return 0;
}
\end{lstlisting}

\ifnotabridged{}
\begin{lstlisting}[float=t, style=CStyle, label={lst:attributes}, caption={Function- and variable- level annotations exposing \WriteBeforeRead \CCs to developers.}]
/* Example function-level annotation */
__attribute__(("writebeforeread")) 
void function_with_potentially_uninitialized_variables() { 
    /* ... */ 
}

/* Example variable-level annotation */
void function_with_potentially_uninitialized_variables() {
    int __writebeforeread uninitialized_variable;
    /* ... */ 
}
\end{lstlisting}
\fi
 
\begin{figure*}[t!]
    \centering
    \begin{subfigure}[t]{0.30\textwidth}
        \centering
        \includegraphics[width=.9\linewidth, scale=0.10]{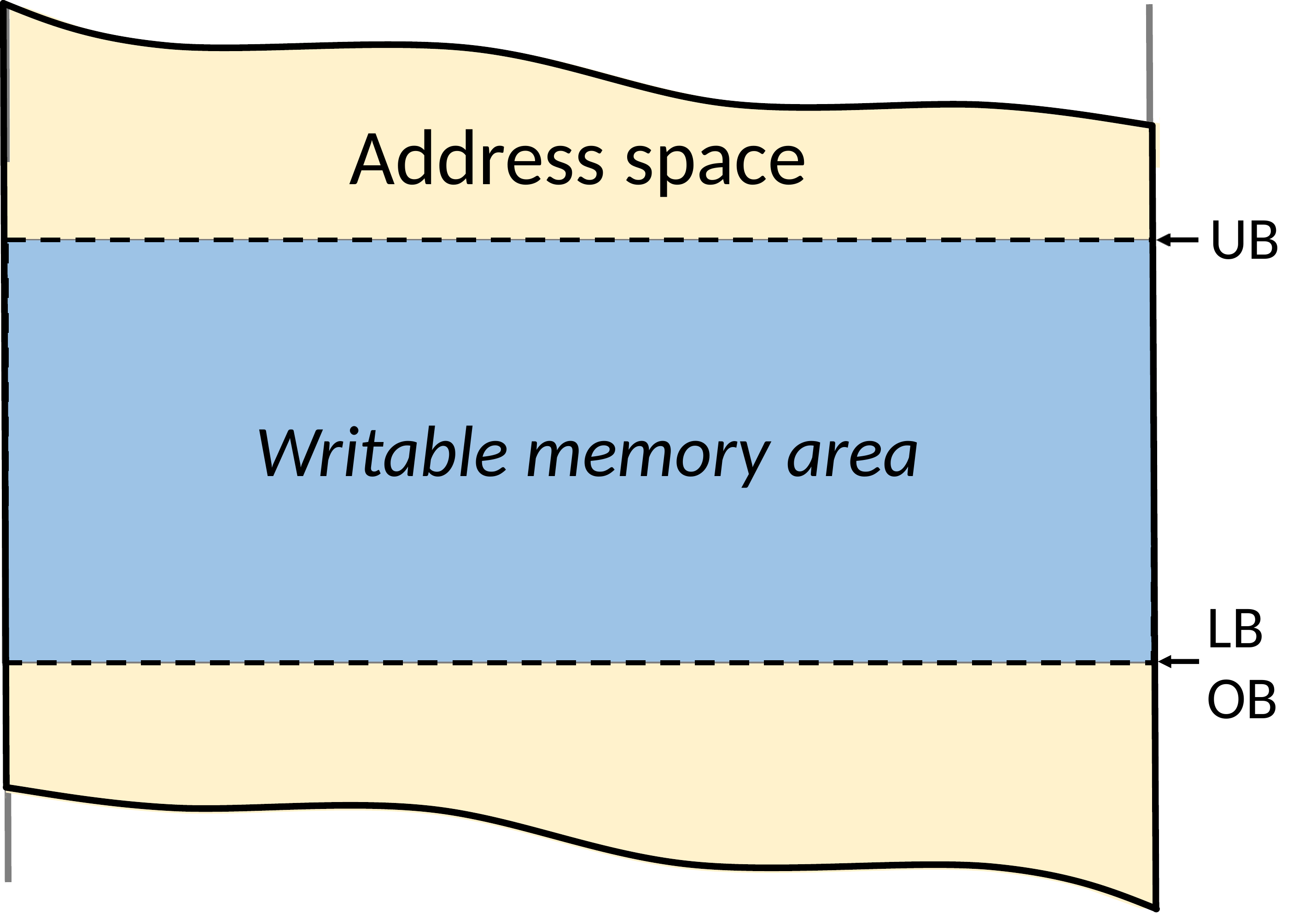}
        \caption{Initial state of memory area.}
    \end{subfigure}
    \begin{subfigure}[t]{0.30\textwidth}
        \includegraphics[width=.9\linewidth, scale=0.10]{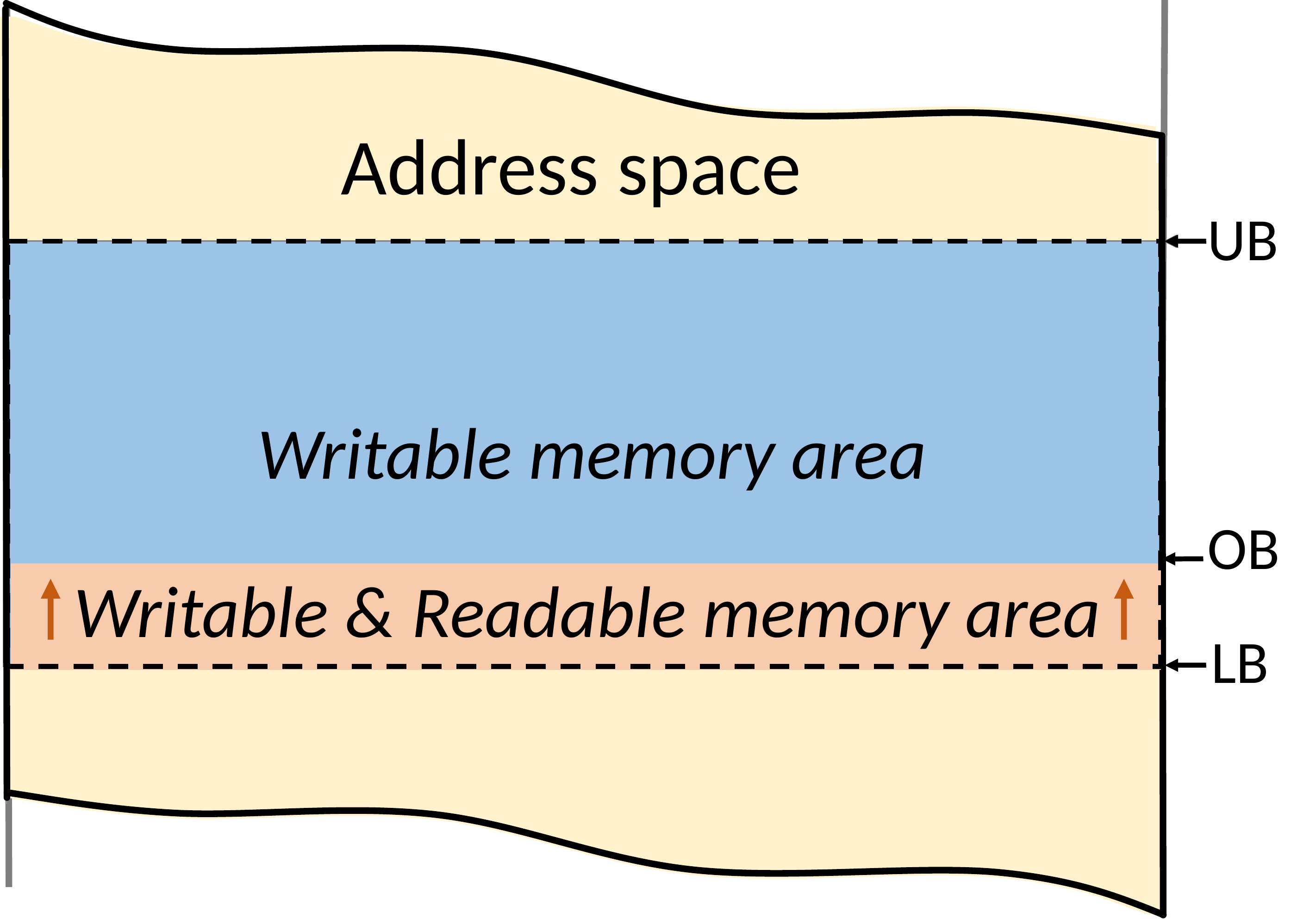}
        \caption{State after store between \acrshort{lowerbound} and \acrshort{opbound}}
    \end{subfigure}
    \begin{subfigure}[t]{0.30\textwidth}
        \includegraphics[width=.9\linewidth, scale=0.10]{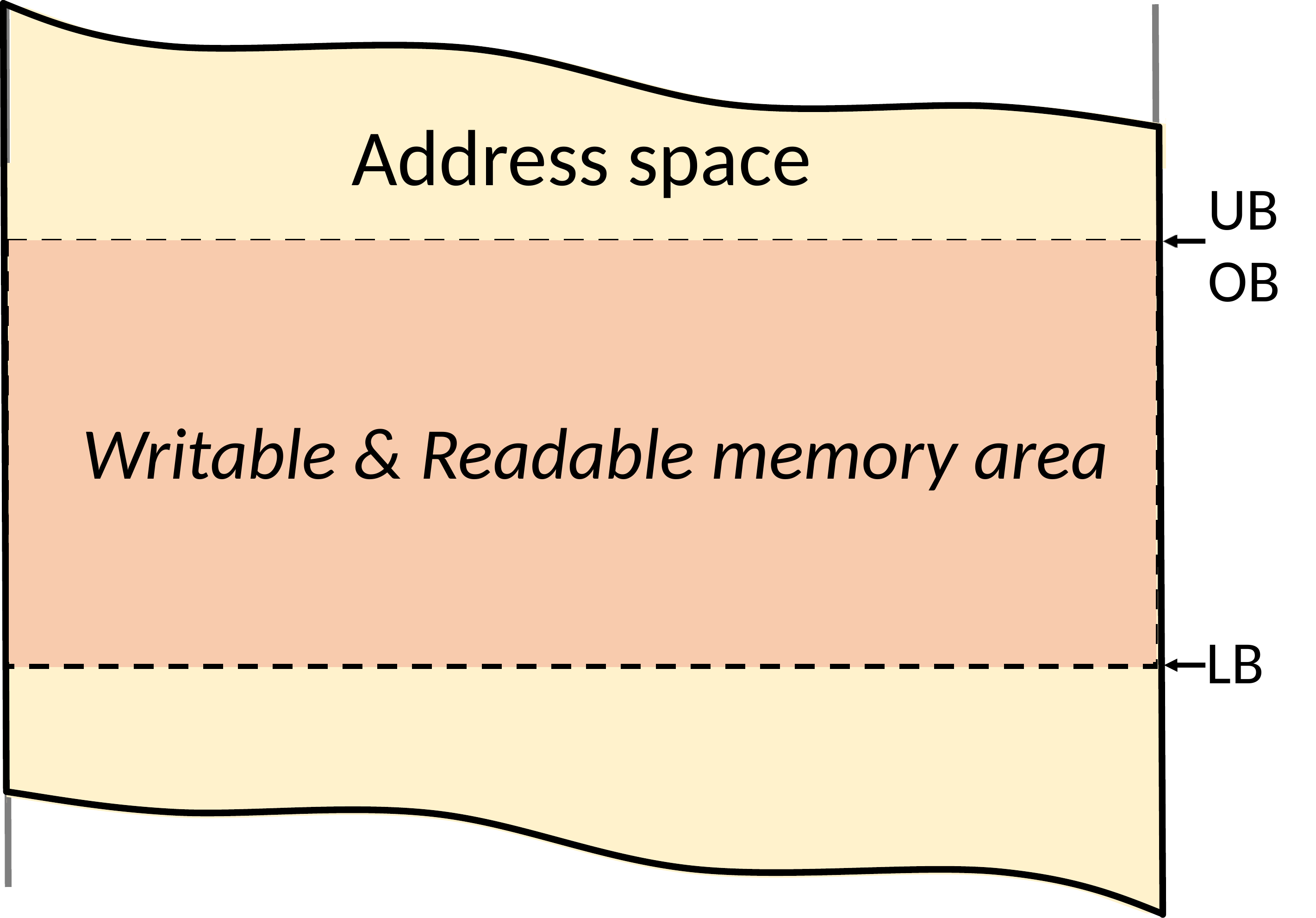}
        \caption{Final state when \acrshort{opbound} reaches \acrshort{upperbound}.}
    \end{subfigure}
    \caption{\WriteBeforeRead{} \CC{} state transitions: \emph{(a)} newly allocated memory under the \WriteBeforeRead{} \CP{} is write-only between the \gls{lowerbound} and \gls{upperbound} and becomes gradually readable \emph{(b)} after store operations advance the \gls{opbound}. In the final state \emph{(c)}, the \gls{opbound} has reached the \gls{upperbound} and the \CC{} behaves identical to a corresponding conventional capability.}\label{figure:highlevelidea}
    %\tn{Replace subfigure (c) with corresponding figure showing the writable + readable area has reached \gls{upperbound}}
\end{figure*}

\begin{table*}[t]
    \caption{\Glsentrydescplural{cp} types for \CCs}\label{table:conditional_permissions}
    \centering
    \begin{tabular}{l p{12cm}}
    \toprule
        \textbf{\Glsentrydescplural{cp}\xspace} & \textbf{Description} \\
    \midrule
    \WriteBeforeRead{} & Memory must be written to at least once before read-access is granted \\
    \WriteBeforeExecute{} & Memory must be written to at least once before execute access is granted \\
    \WriteBeforeReadOnly{} & Memory must be written to exactly once before read access is granted \\
    \WriteBeforeExecuteOnly{} & Memory must be written to exactly once before execute access is granted \\
    \WriteOnce{} & Memory can be written to exactly once \\
\ifnotabridged\ReadOnce{} & Memory can be read exactly once \\
    \ExecuteOnce{} & Memory can be executed exactly once \\
\fi
    \bottomrule
    \end{tabular}
\end{table*}

\begin{figure*}[t]
    \begin{subfigure}[b]{\textwidth}
        \includegraphics[width=\textwidth]{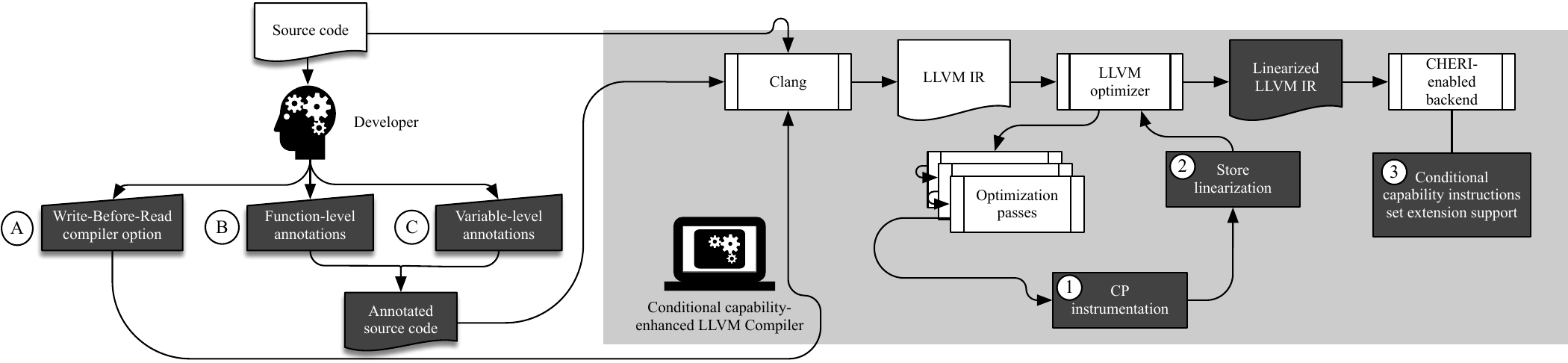}
        \caption{Overview of the \CC{}-enhanced LLVM compiler. Circled letters indicate the different usage models for \CCs: \circled{A} \WriteBeforeRead{} compiler option, \circled{B} function-level annotations, and \circled{C} variable-level annotations. Components with a darker background and circled numbers indicate additions compared to the conventional \gls{cheri}-enhanced LLVM compiler.}\label{fig:arch-compiler}
    \end{subfigure}
    \begin{subfigure}[b]{\textwidth}
        \includegraphics[width=\textwidth]{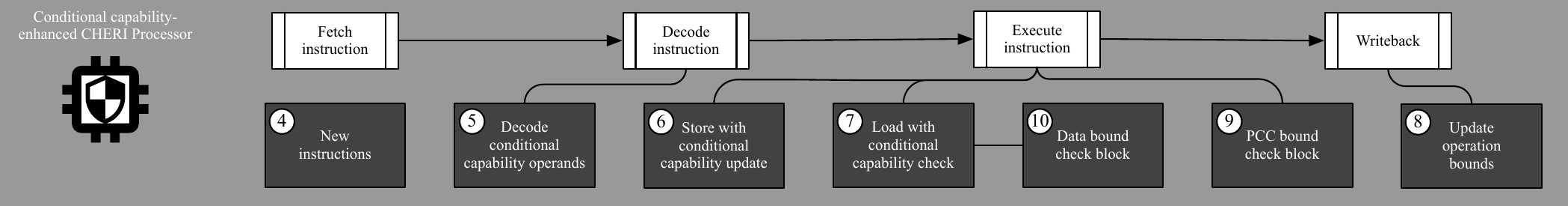}
        \caption{\CC{}-enhanced \gls{cheri} processor. Components with a darker background circled numbers indicate additions compared to the conventional CHERI processor.}\label{fig:arch-cpu}
    \end{subfigure}
    \caption{High-level system architecture of the \CC{}-enhanced LLVM compiler and \gls{cheri} processor.}\label{fig:architecture}
\end{figure*}

\glsunset{opbound}

\subsection{Challenges}\label{sec:challenges}

Integrating \CPs{} into capability-based addressing presents several challenges.
A key issue is managing the tracking of uninitialized data.
Previous approaches~\cite{Georges21,Huyghebaert20,Yu23} assume that data between the lower bound and address pointed to by the capability is initialized, while uninitialized data falls between the address and the upper bound.
However, this approach has limitations, as discussed in~\Cref{sec:relatedwork}.
To address this, we propose \emph{operation-specific bounds} (\glspl{opbound}), which precisely track writes.
\Glspl{opbound} must be updated when a program writes to a region of memory to reflect the new bounds.
Otherwise, stale or misaligned bounds can lead to false positives.

\paragraph{Storing and updating operation bounds.}
To store and update \glspl{opbound} efficiently, we propose leveraging unused bits in the baseline architecture address (\Cref{fig:chericap}).
This minimizes changes to the underlying hardware but makes significant compression of the \gls{opbound} necessary.
Further details on the hardware modifications are provided in~\cref{sec:architecture} and \cref{sec:implementation}.

\paragraph{Integration with existing architectures.}
To be practical, \CPs{} must integrate seamlessly with existing capability architectures, such as \gls{cheri}, that are gaining industry adoption.
As a case study, we integrate \CPs{} to \gls{cheri} while ensuring full compatibility with the underlying RISC-V \gls{isa}.
In \cref{sec:implementation}, we describe the integration process and our hardware extension, ``\emph{\moncheri}'', which introduces only minor modifications to the \gls{cheri} capability representation (\cref{sec:impl-cc}).

\paragraph{Maintaining capability monotonicity.}
\CPs{} add, similar to capability sealing (\Cref{sec:cheri}), controlled non-monotonicity to the \gls{cheri} design.
They allow temporary suspension of permissions, which are regained once the associated condition is met.
Our design (\Cref{sec:cc}) guarantees that \CPs{} do not to elevate privileges beyond the original capabilities.

\paragraph{Maintaining capability linearity.}
\emph{Capability linearity} refers to the consistency and integrity of \glspl{opbound} across a program’s execution.
To maintain this, we propose a new compiler pass to ensure that capabilities are properly updated as the program executes.
This pass is implemented in the LLVM-based \moncheri{} toolchain, as
detailed in \cref{sec:architecture} and \cref{sec:instrumentation}.
%, and ensures \glspl{opbound} are properly updated as the program executes.

\paragraph{Minimizing run-time overhead.}
A concern with any memory access control mechanism is the potential for performance degradation.
For \moncheri{}, we ensure that checks involved in monitoring memory accesses do not add significant latency or computational overhead during normal program execution.
In \Cref{sec:isa-ext}, we explain how the \gls{opbound} checks are carefully designed to minimize additional cycles per instruction.

\subsection{Conditional Capabilities}\label{sec:cc}
% \Glsentrylongpl{cc}

\Glsentrylongpl{cc} are designed to complement the conventional capability permissions with \glsentrylongpl{cp}, allowing permissions to be tailored to specific needs or requirements for different sections of data or operations.
The conventional capability permissions and \CPs, when enabled for a capability, are evaluated in parallel; both sets of permissions need to be valid for memory access to be allowed.
\Glsentrylongpl{cc} consist of two building blocks:
\begin{inparaenum}[1)]
    \item \emph{\Glsentrylongpl{cp}} enable capabilities to trace whether the corresponding condition is fulfilled. 
    \item \emph{operation-specific bounds} enable \CPs to be enforced at a granularity that allows the differentiation between accessible and inaccessible memory ranges within the conventional bounds of the capability.
\end{inparaenum}

\newcommand{\Operation}{\ensuremath{\mathcal{X}}\xspace}
\newcommand{\Perm}{\ensuremath{\mathcal{P^U}}\xspace}
\newcommand{\CondPerm}{\ensuremath{\mathcal{P^C}}\xspace}
\newcommand{\EffectivePerm}{\ensuremath{\mathcal{P}}\xspace}

\newcommand{\AnotherOperation}{\ensuremath{\mathcal{Y}}\xspace}
\newcommand{\AnotherCondPerm}{\ensuremath{\mathcal{Q^C}}\xspace}

\glsreset{opbound}

\paragraph{\Glsentrylongpl{cp}.}
To describe \glsentrylongpl{cp}, we denote conventional permissions that remain \emph{unchanged} throughout the lifetime of a capability, as \Perm, operations on memory as \Operation, and \CPs{} as \CondPerm.
A \CondPerm{} is granted on the condition that \Operation{} occurs: $\Operation \implies \CondPerm$ (if \Operation, then \CondPerm).
If \Operation{} does not occur, \CondPerm{} is not granted.
On an architectural level, operations (\Operation) are limited to load, store and execute, while permissions (\Perm{} and \CondPerm) can be read, write and execute.
For simplicity, we will use the terms \texttt{Write}, \texttt{Read}, and \texttt{Execute} when referring to both \Operation{} and \CondPerm{} in \CP{} names.
The effective permissions (\EffectivePerm) are a subset of the conventional and the conditional permissions:

\[
\EffectivePerm = \begin{cases}
        \Perm & if \ \CondPerm = \emptyset  \\
        \Perm \land \left(\Operation \implies \CondPerm\right)  & if \ \CondPerm \neq \emptyset   \\
    \end{cases}   
\]
\ifabridged
\newpage  % Avoid orphan line at end of previous page
\fi
\paragraph{Operation-specific bounds.}
Tracking the memory range for which $\CondPerm:\Operation \implies \CondPerm$ requires \emph{operation-specific} bounds.
Each \gls{opbound} tracks the subset of memory within a capability's conventional bounds for which \Operation has occurred.
To accommodate multiple \CPs{} at the same time, $\Operation \implies \CondPerm$ and $\AnotherOperation \implies \AnotherCondPerm$ where $\Operation \neq \AnotherOperation$ requires two distinct \glspl{opbound}.
Conversely, for $\Operation \implies \CondPerm$ and $\Operation \implies \AnotherCondPerm$, where $\CondPerm \neq \AnotherCondPerm$, one \gls{opbound} is sufficient.

\Cref{figure:highlevelidea} illustrates an example use case for our conditional capabilities: \WriteBeforeRead.
In its initial state (\Cref{figure:highlevelidea}.a), the capability refers to a writable memory area with upper and lower bounds.
If a store occurs in the memory area (\Cref{figure:highlevelidea}.b), the \gls{opbound} increases by the size of the store operand on the hardware level.

An advantage of our design is that it allows capability hardware to
enforce a variety of novel access-control permissions beyond conventional
\Read{}\Write{}\Execute{}.
We identify the \CPs{} and corresponding use cases in
\Cref{table:conditional_permissions}.
\ifnotabridged{}
This set of \CPs{} allows the
definition of “single-use” code and data that enable policies that can be
used to harden software against run-time attacks that aim to reuse
existing program resource. Such attacks, e.g., code-reuse attacks such as
return-oriented-programming~\cite{Shacham07}, can use pre-existing sequences of instructions
to craft malicious control flows that can be used to compromise program
behavior.  Code that a process executes only once, e.g., a fixed sequence
of initialization instructions, could, using the \ExecuteOnce{} \CP{} be
invoked once during program initialization and is subsequently rendered
useless for code-reuse attacks later in the process’ lifetime.

\fi
\WriteBeforeExecute{} is useful for \gls{jit} compilers that reuse memory buffers for generated code, making them targets for control-flow hijacking attacks using executable heap or stack buffers.
\WriteBeforeExecuteOnly{} enables \CCs{} to emulate \gls{xom}, typically limited to firmware, to protect secrets such as stack canary reference values or values used for control-flow enforcement~\cite{Denis-Courmont20} from being read by attackers.

\subsection{High-Level System Architecture}\label{sec:architecture}

\Cref{fig:architecture} depicts the high-level system architecture for our prototype \CC{}-enhanced LLVM compiler and \gls{cheri} processor.
Inputs and process blocks in a darker color indicate the changes to a conventional \gls{cheri}-enabled compiler and processor architecture.
For conciseness, in this description, we focus on \WriteBeforeRead{} \CPs{}.

In our prototype, \CCs{} are exposed to developers through the three alternative usage models: \circled{A} a \WriteBeforeRead{} compiler option, \circled{B} function-level annotations, and \circled{C} variable-level annotations.
Our current design of the compiler option \circled{A} applies \WriteBeforeRead{} to all stack variables, but we discuss possibilities for more intelligent heuristics in \Cref{sec:discussion}.
Function- and variable-level annotations are implemented as Clang function and variable attributes, respectively\ifnotabridged{} (shown in \Cref{lst:attributes})\fi, which gives developers control over which variables \WriteBeforeRead{} is applied to.

The \CC{}-enhanced LLVM compiler (\Cref{fig:arch-compiler}) can operate either on original source code or source code annotated by the developer using the aforementioned attributes. 
This differs from the conventional \gls{cheri}-enabled LLVM in three ways: \dCOne{} the \emph{\CP{} instrumentation} transform pass that marks variable for instrumentation based on the compiler option or developer annotations, 
and emits intrinsics that interface with the \CC{} hardware to initialize operation-specific bounds for newly created capabilities, 
\dCTwo{} a \emph{store linearization} transform pass that ensures that accesses to uninitialized objects in memory are performed consistently through a capability that tracks the operation bound, 
and \dCThree{} support for the new \CC{} instructions in the CHERI RISC-V back end, allowing the compiler to interface with the \CC{}-enhanced \gls{cheri} processor.

The \CC{}-enhanced CHERI processor (\Cref{fig:arch-compiler}) is modified to support the \CC{} instructions (\dCFour, \Cref{sec:isa-ext}), 
and its instruction pipeline is augmented with logic to encode and decode capabilities containing an additional operation bound (\dCFive, \Cref{sec:cc}). 
Modified \dCSix{} store and \dCSeven{} load logic updates and checks the operation bounds when executing store and load instructions, respectively.
Finally, \dCEight{} the pipeline performs writebacks of updated capabilities in register operands after the operations that update the operation bounds.
Code capability checks for the \gls{pcc} are done in a dedicated bounds check block \dCNine{} (needed for \WriteBeforeExecute, \WriteBeforeExecuteOnly\ifnotabridged{}, and \ExecuteOnce)\fi, while all data \CPs{} use a general-purpose bounds check block \dCTen. 

\section{\moncheri{} Implementation}\label{sec:implementation}
\begin{table*}[t!]
    \centering
    \caption{Architectural changes to CHERI-RISC-V by \glsentrydesc{cp}. The \textbf{CSetOpBounds variants} column shows the corresponding \texttt{CSetOpBounds} instruction mnemonic for each \gls{cp}. The \textbf{Pipeline changes} column shows the impact on the \textit{Load}, \textit{Store}, and \textit{Execute} logic inside the processor. A \Circle{} indicates the operation is unaffected, a \LEFTcircle{} indicates the operation is augmented with an \gls{opbound} check, a \RIGHTcircle{} indicates the operation is augmented with a writeback that, under certain conditions, updates the \CC{} \gls{opbound}. Finally, a \CIRCLE{} indicates the operation is augmented with both.}\label{table:instructions}
    \setlength{\tabcolsep}{6pt} % Smaller padding for the first row
    \begin{tabular}{@{}lcccc@{}}\toprule
        \multicolumn{1}{c}{\multirow{2}{*}{\shortstack[c]{\textbf{Conditional}\\\textbf{Permission}}}} &
        \multirow{2}{*}{\shortstack[c]{\textbf{CSetOpBounds}\\\textbf{variants}}} &
        \multicolumn{3}{c}{\textbf{Pipeline changes}} \\
        \cmidrule(lr){3-5}
        \setlength{\tabcolsep}{10pt}  % Increased padding for following rows
        &  & \textit{Load} & \textit{Store} & \textit{Execute} \\ 
        \midrule
        \WriteBeforeRead{}          & \texttt{csetwbrbound} & \LEFTcircle{} & \RIGHTcircle{} & \Circle{}     \\[0.2em]  % Adding space between rows after the first one
        \WriteBeforeExecute{}       & \texttt{csetwbxbound} & \Circle{}     & \RIGHTcircle{} & \LEFTcircle{} \\[0.2em]
        \WriteBeforeReadOnly{}      & \texttt{csetrobound}  & \LEFTcircle{} & \CIRCLE{}      & \Circle{}     \\[0.2em]
        \WriteBeforeExecuteOnly{}   & \texttt{csetxobound}  & \Circle{}     & \CIRCLE{}      & \LEFTcircle{} \\[0.2em]
        \WriteOnce{}                & \texttt{csetwtbound}  & \Circle{}     & \CIRCLE{}      & \Circle{}     \\[0.2em]
\ifnotabridged\ReadOnce{}                 & \texttt{csetrtbound}  & \CIRCLE{}     & \RIGHTcircle{} & \Circle{}     \\[0.2em]
        \ExecuteOnce{}              & \texttt{csetxtbound}  & \Circle{}     & \Circle{}      & \CIRCLE{}     \\ 
\fi
        \bottomrule
    \end{tabular}
\end{table*}

In this section, we present \moncheri{}, our implementation of \CPs{} for the CHERI-RISC-V architecture. 
We implemented two versions of \moncheri{}: 
\begin{inparaenum}
\item a \moncheri{} software model for the QEMU-system-CHERI128 full-system emulator, and 
\item a \moncheri{} softcore based on the \CheriFlute{} processor \gls{ip}.
\end{inparaenum}
The \moncheri{} \gls{isa} extension is described in \Cref{sec:isa-ext}.

In addition, we extended the existing CHERI-LLVM toolchain~\cite{CTSRD24}
with support for the \moncheri{} \gls{isa} extension and \WriteBeforeRead{}
\CP{} instrumentation for stack-allocated variables (\Cref{sec:instrumentation}). 
%Our changes to CHERI-LLVM are described in \Cref{sec:instrumentation}.
Finally, we added run-time \WriteBeforeRead{} \CP{} support for
heap-allocated memory in the \CheriFreeRtos~\cite{Almatary22} and
\gls{tlsf}~\cite{Conte16} memory allocators (\Cref{sec:runtime}).
%Our changes to the allocators are described in \Cref{sec:runtime}.

\newsavebox{\chericompressed}
\begin{lrbox}{\chericompressed}
\begin{bytefield}[bitwidth=1em]{64}
    \bitheader[endianness=big]{0,2,3,13,14,16,17,25,26,27,44,45,46,47,48,59,60,63} \\
    \bitbox{4}{\CCSWPerms} & \bitbox{12}{\CCHWPerms} & \bitbox{1}{\CCFlags} & \colorbitbox{lightgray}{2}{} &
    \bitbox{18}{\CCOType} & \bitbox{1}{\CCInternalExponent} &
    \bitbox{9}{\CCExpNonZeroTop} & \bitbox{3}{\CCExpHighPart} &
    \bitbox{11}{\CCExpNonZeroBase} & \bitbox{3}{\CCExpLowPart} \\
    \begin{rightwordgroup}{Cursor}
    \bitbox{64}{\CCAddress}
    \end{rightwordgroup} \\
    \end{bytefield}
\end{lrbox}

\newsavebox{\opboundscompressed}
\begin{lrbox}{\opboundscompressed}
    \begin{bytefield}[bitwidth=1em]{64}
        \bitheader[endianness=big]{0,2,3,13,14,16,17,25,26,27,44,45,46,47,48,59,60,63} \\
        \bitbox{4}{\newtext{\CCOpPerms}} & \bitbox{12}{\CCHWPerms} & \bitbox{1}{\CCFlags} & \colorbitbox{lightgray}{2}{} &
        \bitbox{18}{\CCOType} & \bitbox{1}{\CCInternalExponent} &
        \bitbox{9}{\CCExpNonZeroTop} & \bitbox{3}{\CCExpHighPart} &
        \bitbox{11}{\CCExpNonZeroBase} & \bitbox{3}{\CCExpLowPart} \\
        \begin{rightwordgroup}{Cursor}
        \bitbox{11}{\newtext{\CCExpNonZeroOp}} & \bitbox{5}{\newtext{\CCExtOp}}
        \bitbox{48}{\CCAddress}
        \end{rightwordgroup} \\
    \end{bytefield}
\end{lrbox}

\subsection{\moncheri{} Extension for CHERI-RISC-V}\label{sec:isa-ext}

\paragraph{\texttt{CSetOpBounds} instructions.}
The RISC-V \gls{isa} is extensible through portions of the instruction encoding space reserved for \gls{isa} extensions.
We add a family of \texttt{CSetOpBounds} instructions to CHERI-RISC-V to the existing CHERIv9~\cite{Watson23} non-standard \texttt{Xcheri} extension in the \emph{custom-2/rv128} opcode space to allow setting an initial, or updating an already set, value for a capability's operation bound.
Our \texttt{CSetOpBounds} instructions require two operands: an existing capability and a length operand.
When invoked, \texttt{CSetOpBounds} sets the operation bound to the range $\left[b, o\right]$ where \acrshort{base} is the ``\emph{base}'' encoded as part of the conventional capability bound, and \acrshort{optop} is the ``\emph{operation top}'' encoded into the operation bound (see~\Cref{sec:impl-cc}).
\ifnotabridged{}
\texttt{CSetOpBounds} does not allow programs to increase the operation bounds,  while they are allowed to invoke \texttt{CSetOpBounds} to decrease the operation bounds of a capability. 
\fi
The operational bounds set by \texttt{CSetOpBounds} are restricted to be within the original capability bounds.

Additionally, \texttt{CSetOpBounds} must identify the \CPs{} which operation bound to set.
Lacking free operands in the instruction encodings available in \texttt{Xcheri}, we chose to encode the \CP{} information into the \texttt{CSetOpBounds} opcode.
The \texttt{CSetOpBounds} instruction variants are shown in Table~\ref{table:instructions}.

\paragraph{\Glsentrylongpl{cp}.}
We assign new \emph{\glsentrylongpl{cp} control bits} from the \gls{cheri} capability representation (\CCOpPerms{} in \Cref{fig:opboundscompressed}).
The \CheriFlute{} \gls{ip} reserves 12-bits for hardware-defined permissions, i.e., those authorizing load, store, etc., and 4-bits for software-defined permissions, which interpretation are left open by the CHERIv9 specification.
For the purposes of prototyping \CPs{}, we utilize the four available software-defined bits, treating them as a 4-bit integer that enumerates one of 16 possible, mutually exclusive permission states.
Five of those states correspond to the \CPs{} in \cref{table:instructions}, ten modes are left unused, and one mode, the zero value, corresponds to the default state in which \CP{} enforcement is disabled.
In this default state, protections such as \WriteBeforeRead{} are inactive, allowing uninitialized memory access within the capability's bounds. 
Invoking \texttt{csetwbropbound} on a capability enables \WriteBeforeRead{} enforcement for that capability.

\paragraph{Processor pipeline changes.}
To avoid increasing processor latency, bounds checks are carefully structured within the execute instruction block of the pipeline (see Figure~\ref{fig:arch-cpu}), which has two stages.
In the first stage, where single-cycle \gls{alu} operations are performed, the \gls{cheri} implementation checks the conditional permissions and initiates the data capability bounds check, while the actual bounds check occurs in the second stage (used for longer-latency operations like memory access).
\Gls{opbound} checks run in parallel with the general-purpose bounds checking to minimize their latency.
For stores and loads, the first stage checks if the conditional permission allows the operation, and if the target address results in an update of the current operation bound; if it does and is adjacent, the second stage extends the bound to cover it, with the update committed in the writeback stage.
% For loads, and instruction fetches, the first stage ensures the address is within the operation bounds.
Table~\ref{table:instructions} shows how the \gls{opbound} checks and writeback are used by different \CPs. 
Any violation raises a \gls{cheri} protection exception in the second stage.

\ifabridged{}
\begin{figure*}
    \resizebox{\textwidth}{!}{\usebox{\opboundscompressed}}
    \caption{Layout of a 128-bit \moncheri{} capability with compressed operation bound. The labels indicate the fields for \CP{} control bits (\CCOpPerms), hardware-defined permissions (\CCHWPerms{}), flag (\CCFlags), object type (\CCOType), internal exponent (\CCInternalExponent), compressed top (\CCTop) including high part of exponent (\CCExpHighPart), compressed base (\CCBase) including low part of exponent (\CCExpLowPart), and cursor consisting of the compressed operation bound (\CCOp, \CCExtOp) and pointer address (\CCAddress). Adapted from~\cite{Watson23} with changes compared to CHERI ISAv9 compressed capability representation \newtext{marked in red.}}\label{fig:opboundscompressed}
\end{figure*}
\else
\begin{figure*}
    \begin{subfigure}{\textwidth}
        \resizebox{\textwidth}{!}{\usebox{\chericompressed}}
        \vspace{-12pt}
        \caption{128-bit CHERI Capability with CHERI Concentrate compression (adapted from~\cite{Watson23})}\label{fig:chericompressed}
    \end{subfigure}
    \begin{subfigure}{\textwidth}
        \resizebox{\textwidth}{!}{\usebox{\opboundscompressed}}
        \caption{128-bit CHERI Capability with compressed operation bound}\label{fig:opboundscompressed}
    \end{subfigure}
    
    \caption{Layouts of 128-bit CHERI Capabilities without (\subref{fig:chericompressed}) and with (\subref{fig:opboundscompressed}) compressed operation bound. The labels indicate the fields for software-defined and hardware-defined permissions (\CCSWPerms{} and \CCHWPerms{} respectively), \CP{} control bits (\CCOpPerms), flag (\CCFlags), object type (\CCOType), internal exponent (\CCInternalExponent), compressed top (\CCTop) including high part of exponent (\CCExpHighPart), compressed base (\CCBase) including low part of exponent (\CCExpLowPart), and cursor consisting of the pointer address (\CCAddress) in \subref{fig:chericompressed}, the address and compressed operation bound (\CCOp, \CCExtOp) in \subref{fig:opboundscompressed}.}\label{fig:compression}
\end{figure*}
\fi

\paragraph{Avoiding data hazards.}
Conditional Capabilities create dependencies between instructions that are typically independent.
This is because stores using \CCs{} update the \gls{opbound} in the capability operand, while stores on conventional capabilities do not.
As a result, when a store using a \CC{} is immediately followed by a load or another store using the same capability, a \emph{data hazard occurs}: the latter operation requires the updated \gls{opbound} at the \gls{alu} stage for the \gls{opbound} check but is not available in the operand register until the writeback stage completes. 

To solve this, we implemented a \emph{bypass}---a new data path within the pipeline---that forwards the updated \gls{opbound} from the memory access or writeback stage to the \gls{alu} stage.
This bypass activates when the processor detects that the next instruction in the pipeline operates on the same \CC{}.
This avoids the need to stall the processor or reorder instructions during
compilation and is compatible with conventional RISC-V instruction
ordering. Appendix~\Cref{app:datahazard} shows an example of a data hazard
that is avoided by the bypass.

\subsection{Adding Operation Bounds to Capabilities}\label{sec:impl-cc}

The \gls{cheri} \gls{isa} introduced capability compression in ISAv6~\cite{Watson17}, with the current \gls{cheri} Concentrate\cite{Woodruff19} compression scheme introduced in ISAv7~\cite{Watson19a}.
Compression reduces the in-memory size of capabilities, which would otherwise occupy 256 bits, quadrupling the space needed for native 64-bit pointers, and increasing cache footprint and memory bandwidth requirements.
Expanding capabilities further to include additional operation bounds is impractical.
However, most modern 64-bit \glspl{os} use only part of the 64-bit virtual address space.
For instance, RISC-V Linux uses 48-bit addresses~\cite{Ghiti21}, and 32-bit \glspl{os} like FreeRTOS use 32 bits even on 64-bit hardware.
We leverage these unused bits to store an additional 16-bit \gls{opbound}.

\paragraph{\Glsentrylong{cc} masking.}
Masking a portion of the address in cases where \glspl{os} do not fully utilize 64-bit addresses has precedents in both conventional processors, such as in Arm's \gls{tbi}~\cite{Spickett23}, Intel \gls{lam}~\cite{Intel24}, and AMD \gls{uai}~\cite{AMD24} features, as well in the \gls{cheri} design as alternative compression formats described in Appendix~E of the \gls{cheri} ISAv9~\cite{Watson23}.
To encode operation bounds, \moncheri{} applies an address mask when a register holds a \CC{}.
Since CHERI-RISC-V uses a merged register file, general-purpose registers can hold either a 64-bit integer or a 128-bit capability. 
Unlike prior RISC-V pointer masking proposals~\cite{Maas24}, \moncheri{}'s address mask is limited to \CCs, leaving conventional \gls{cheri} capabilities and regular pointers unchanged.

\ifnotabridged{}
\begin{figure*}
\noindent\begin{minipage}{\textwidth}
\begin{lstlisting}[style=CStyle, label={lst:ir}, caption={LLVM \gls{ir} of \Cref{lst:attributes}} after \CP LLVM Pass Instrumentation]
define dso_local void @function_with_potentially_uninitialized_variables() local_unnamed_addr addrspace(200) #11 !dbg !322 {
entry:
%*\dWdth*)%uninitialized_variable = alloca i32, align 4, addrspace(200), !clang.decl.ptr !325, !clang.var.writebeforeread !100
%*\dOne*)%0 = call ptr addrspace(200) @llvm.cheri.bounded.stack.cap.i64(ptr addrspace(200) %uninitialized_variable, i64 4)
%*\dTwo*)%1 = call ptr addrspace(200) @llvm.cheri.cap.op.bounds.set.i64(ptr addrspace(200) %0, i64 0)
 %*\dWdth*)/* ... */
}
\end{lstlisting}
\end{minipage}
\end{figure*}
\fi

\ifabridged{}
\paragraph{\moncheri{} compressed capability representation.}
\else
\paragraph{\gls{cheri} compressed capability representation.}
\fi
A raw 256-bit capability is comprised of three virtual addresses: \gls{base}, \gls{top}, and \gls{address}.
The 128-bit representation of capabilities utilizes the redundancy between the three addresses and stronger alignment requirements (proportional to object size) for a more compact representation.

\ifabridged{}
\Cref{fig:opboundscompressed} shows the capability format for \moncheri.
\else
\Cref{fig:chericompressed} shows the capability format for \gls{cheri} Concentrate.
\fi
\CCBase{} and \CCTop{} encode the \gls{base} and \gls{top} bounds in one of two formats depending on the \gls{ie} bit: if $\CCInternalExponent = 1$ then an \gls{exponent} is stored in the lower three bits of \CCBase{} and \CCTop{} (\CCExpLowPart{} and \CCExpHighPart) reducing their precision by three bits.
\CCExponent{} determines the position at which \CCBase{} and \CCTop{} are inserted into \CCAddress{} to obtain \gls{base} and \gls{top}.
Otherwise ($\CCInternalExponent = 0$, $\CCExponent = 0$) the full width of \gls{base} and \gls{top} are used.
Their width is determined by an encoding parameter: \gls{mw} that determines the precision of the decoded bounds.
The \gls{cheri} ISAv9 uses $\CCMantissaWidth = 14$ for 128-bit capabilities.
However, \gls{top} is further compressed by two bits as the top two bits of \gls{top} can be derived from the equation $\CCTop = \CCBase + L$ where the most significant bit of $L$ (\LMsb) is known from the values of \CCInternalExponent{} and \CCExponent{} and a carry bit is implied if $\CCExpZeroTop < \CCExpZeroBaseBitsToCompare$ since \gls{top} is known to be larger than \gls{base}.

When decoding the bounds, \gls{base} and \gls{top} are derived from \gls{address} by substituting \gls{mw} bits, \CCExponent{} to $\CCExponent + \CCMantissaWidth$, with \CCBase{} and \CCTop{} and clearing the bottom \CCExponent{} bits.
To allow \CCAddress{} to span a larger region while maintaining the original bounds, the most significant bits of \gls{top} and \gls{base} $ a_{top} = a\left[63 : \CCExponent + \CCMantissaWidth\right]$ can be adjusted up or down using corrections $c_t$ and $c_b$.
The detailed description of the \gls{cheri} Concentrate compression can be found in Section 3.5.4 of the \gls{cheri} ISAv9 specification~\cite{Watson23}.

\ifnotabridged{}
\paragraph{\moncheri{} compressed capability representation.}
\Cref{fig:opboundscompressed} shows the capability format for \moncheri.
\else
\paragraph{Encoding the operation bound.} 
\fi
We encode the operation bound, \gls{optop} in two fields in the most significant 16-bits of the cursor: \CCExpNonZeroOp{} (11 bits) and \CCExtOp (5 bits) which are freed by limiting $ a_{top} = a\left[47 : \CCExponent + \CCMantissaWidth\right]$.
When $\CCInternalExponent = 0$, \gls{optop} is stored identically to \gls{base} with its lowest three bits \CCBottomBitsOp derived from the most significant bits of \CCExtOp{} (\CCExpZeroExtOpHighPart).
When $\CCInternalExponent = 1$, up to five bits from $\CCExtOp\left[E+2:0\right]$ are used to store the least significant bits of \gls{optop}.
Our current implementation limits \CCExtOp{} to five bits, limiting \gls{optop} to an $\CCInternalExponent$ of at most $2$.
We discuss methods to alleviate this limitation in \Cref{sec:discussion}.
Expanding \gls{optop} to full 48 bits happens similarly to \gls{base} with its own correction $c_o$.
\Cref{fig:format} in \Cref{app:decoding} illustrates the changes relative to \gls{cheri} Concentrate compression.

\subsection{\moncheri{} Support for CHERI-LLVM}\label{sec:instrumentation}

\paragraph{\CP{} instrumentation.}
\ifnotabridged{}
\Cref{lst:ir} shows an excerpt of the LLVM \gls{ir} of one of the functions in \Cref{lst:attributes} after \CP{} instrumentation (see \dCOne{} in \Cref{fig:architecture}). 
\fi
The CHERI-LLVM compiler adds capability bounds to stack variables in an \gls{ir}-level compiler pass, \texttt{CheriBoundAllocas}.
This pass replaces every \gls{ir} stack allocation instruction (\texttt{alloca}\ifnotabridged{}, \dOne{} in \Cref{lst:ir}\fi) with a \longword{llvm.cheri.cap.bounds.set} intrinsic.
These are then replaced with \gls{cheri} \texttt{csetbounds} instructions by the CHERI-RISC-V backend.
We extended this pass to add \longword{llvm.cheri.cap.op.bounds.set} intrinsics~\ifnotabridged{} (\dTwo{} in\Cref{lst:ir})\fi for variables that are either annotated with \texttt{\_\_writebeforeread} (\circled{C} in \cref{fig:arch-compiler}), in functions annotated with  \texttt{\_\_attribute\_\_(("writebeforeread"))} (\circled{B}), or all stack variables, when compiling with the \CheriWriteBeforeRead{} option (\circled{A}).
The \longword{llvm.cheri.cap.op.bounds.set} intrinsic is replaced with \texttt{csetwbrbound} instructions by the backend. 

\paragraph{Optimizing \WriteBeforeRead{}.}
The \texttt{CheriBoundAllocas} pass checks if \texttt{alloca} instructions fall within the original capability bounds and omits \longword{llvm.cheri.cap.bounds.set} in those cases.
We disable this optimization for variables that receive \longword{llvm.cheri.cap.op.bounds.set} since it does not guarantee stores to \WriteBeforeRead{} variables fall within the \glspl{opbound}.
However, not all variables require run-time \CP{} checks. 
For \WriteBeforeRead{}, we optimize arrays and scalar variables allocated in function entry blocks by using a simple, non-heuristic analysis that checks if they are fully initialized before being accessed. 
The analysis inspects each store in the function's first basic block and verifies whether they are preceded by loads to the same allocation.
For arrays, we track initialization with a vector, checking store instructions for the base pointer or specific indices, ensuring each load is preceded by a store through dominance analysis\cite{Lowry69}; 
if a load occurs before a store, the index is marked uninitialized in the vector.
For scalars, we verify whether a store targets the variable, and each load is dominated by a corresponding store.
Variables shown to be initialized do not receive the \WriteBeforeRead{} \CP{} and, therefore, can be checked by \texttt{CheriBoundAllocas} for capability-bound optimization.
In \Cref{sec:discussion}, we discuss the possibility of using heuristic static analysis to optimize \WriteBeforeRead{} variables further.

\paragraph{Store linearization.}
LLVM \gls{ir} uses \gls{ssa} form, where each variable has exactly one assignment.
The compiler creates new \gls{ir} variables to maintain this rule when a variable has multiple assignments.
If a variable is accessed through different control-flow paths, a phi ($\Phi$) function is introduced to merge the values from these paths.

During register allocation, the compiler maps variables to processor registers. However, due to register pressure---the number of live variables exceeding the number of available registers—--the compiler generates \emph{spill code} that moves variable contents between memory and registers.
Live-range splitting optimizes when variables are spilled and can make use of the fact that the same variable may, at times, be available in multiple registers simultaneously.
However, when a \CC{} is stored in multiple registers, the state of its operation bound may become inconsistent across instances.
This can occur due to:
\begin{inparaenum}
\item register spilling, or
\item register forking, where the \CC{} state differs between registers.
\end{inparaenum}
Inconsistencies arise when a duplicated capability is stored in memory, becomes stale as its copy's operation bound is updated, and is later restored.

To address this, we introduce a \emph{store linearization pass}, which runs after \gls{ssa} optimizations and ensures a single, canonical \CC{} instance is maintained across stores. 
Performing store linearization after optimizations guarantees robustness across optimization levels.

Store linearization inserts placeholder function calls for every store operand in the \gls{ir} to prevent live-range splitting from spilling a canonical \CC, causing it to grow stale.
This placeholder is an identity function---effectively a no-op---taking a \CC{} operand and returning it unchanged.
This, however, creates a data dependency between the input and output operand, signaling to the live-range splitting algorithm that the \CC{} has changed since its use in the store.

Array indexing via the LLVM \gls{ir} \gls{gep} instruction requires special handling.
\gls{gep} takes a pointer and an offset, returning a new pointer (in \gls{ssa}-form) pointing at the specified offset in the array.
Store linearization inserts a placeholder function for the \gls{gep} operand, not the indexing pointer.
In practice, the CHERI-RISC-V backend generates code using integer-relative store instructions (\texttt{s[bhwd]}), where the \gls{gep} input is used as a capability operand with an offset.
Therefore, the linearization targets the original operand to extend its liveness.

Finally, the pass recursively replaces any variables holding \CCs{} used store operands with those returned by the corresponding placeholder functions.

\paragraph{Escape value analysis.}
When a value in \gls{ssa} form escapes a code block, it must be updated when accessed along different control-flow paths.
The \texttt{reg2mem} transformation~\cite{LLVMteam19} in LLVM handles escaped values by:
\begin{inparaenum}[1)]
    \item allocating stack memory for each escaped \gls{ssa} register, 
    \item storing \gls{ssa} values in this memory before exiting a block, and 
    \item reloading values upon entering a new block.
\end{inparaenum}
This ensures consistency across code paths.
Store linearization modifies \texttt{reg2mem} to update \CCs{} across control-flow paths, similar to how phi functions merge \gls{ssa} values.
One optimization available to store linearization stems from the original \texttt{reg2mem} also tracking output operands from the \gls{gep} instruction as escaped values.
The special handling of array indexing allows us to avoid storing output operands from \gls{gep} in memory.

Similarly, \CCs{} stored explicitly in memory must be updated after being used for store operations, even in straight-line code.
Store linearization recursively checks if the operand used in a store is itself stored in memory.
If so, an additional store operation is inserted to update the capability stored in memory.
Listings~\ref{lst:64bsink} and~\ref{lst:64bsinkir} in \Cref{app:juliet} illustrate this transformation.

\subsection{\moncheri{} Support for Memory Allocators}\label{sec:runtime}
\ifnotabridged
\begin{lstlisting}[float=t, style=CStyle, label={lst:heap}, caption={Example of \CP-enhanced version of \texttt{malloc().}}]
#include <cheriintrin.h>

void *ptr malloc (size_t size) {
   %*\dWdth*)/*Dynamic Memory Memory management */
   %*\dCOne*)cheri_bounds_set(ptr, size);
   %*\dCTwo*)cheri_opbounds_set(ptr, 0, WriteBeforeRead);
   %*\dWdth*)return ptr;
}
\end{lstlisting}
\fi
We implemented the \texttt{cheri\_opbounds\_set()} \gls{api} to provide low-level support for \CCs{} in system software, such as memory allocators.
To enable \WriteBeforeRead{} \CPs{} for heap allocations, the allocator must initialize the capabilities to allocate memory with the appropriate \CP{}.

\ifnotabridged{}
\Cref{lst:heap} shows how we modified the \CheriFreeRtos{} memory allocator, and the \gls{tlsf} allocator~\cite{Conte16}, previously ported to \gls{cheri}~\cite{Ruchlejmer24} to enforce \WriteBeforeRead{} on allocation made by \texttt{malloc()}.
\fi{}
A \gls{cheri}-aware \texttt{malloc()} already uses the \texttt{cheri\_bounds\_set()} intrinsic to set bounds of the return pointer, \texttt{ptr}, based on the allocation size.
To make it \CC{}-aware, we added a call to \texttt{cheri\_opbounds\_set(ptr, 0, WriteBeforeRead)} before returning from \texttt{malloc()}.
This explicitly sets the operation bounds of \texttt{ptr} to zero and configures the \CP{} control bits to indicate \texttt{ptr} should be treated as \WriteBeforeRead{} by the hardware.

\section{Evaluation}\label{sec:evaluation}
\newcommand*\truepositive{\cellcolor{green!25} \footnotesize}
\newcommand*\truenegative{\cellcolor{green!25} \footnotesize}
\newcommand*\falsepositive{\cellcolor{red!25} \footnotesize}
\newcommand*\falsenegative{\cellcolor{red!25} \footnotesize}

\begin{table}[t!]
    \centering
    \caption{Detection rate of \moncheri on uninitialized memory issues from Juliet Test Suite~\cite{nsajuliet} CWE457 test cases. Green cells indicate the true positives and true negatives, while red cells indicate the false positives and false negatives.}
    \label{tab:rwtable}
    \begin{tabular}{c c | c}\toprule
      \multirow{2}{*}{Ground truth}  & \multicolumn{2}{c}{\moncheri}                              \\
                                     &  \multicolumn{1}{c}{\textbf{Positive}} & \textbf{Negative} \\ 
      \cmidrule(lr){2-3}
      \multirow{2}{*}{\textbf{Bad}}  & \cellcolor{green!25} 560 &  \cellcolor{red!25}   0         \\ 
                                     & \truepositive{100\%}     &  \falsenegative{0\%}            \\ \hline
      \multirow{2}{*}{\textbf{Good}} & \cellcolor{red!25} 6$^*$ &  \cellcolor{green!25} 554       \\
                                     & \falsepositive{1\%}      &  \truenegative{99\%}            \\ \bottomrule
    \end{tabular}
    \\ \parbox{\dimexpr \columnwidth-1em}{$^*$\footnotesize Following the ``Good'' and ``Bad'' classification in Juliet, \moncheri reports six false-positives. Yet, these cases do exhibit uninitialized memory access behavior (cf.~\Cref{sec:securityeval}).}
\end{table}

\begin{table*}[t!]
    \centering
    \caption{Area cost on VCU118 @ 100MHz expressed in number of \glspl{lut} and number of registers.}\label{tab:hwcost}
    \begin{tabular}{r rcc rcc rcc}\toprule
        & \multicolumn{6}{c}{\textbf{\glspl{lut}}} & \multicolumn{3}{c}{\textbf{registers}} \\
        \cmidrule(lr){2-7}\cmidrule(lr){8-10}
                                     & logic & \multicolumn{2}{c}{$\Delta$} &  memory & \multicolumn{2}{c}{$\Delta$} & registers & \multicolumn{2}{c}{$\Delta$} \\ \midrule
        \CheriFlute                  & 139109 & \multicolumn{2}{c}{--}       &    10649 & \multicolumn{2}{c}{--}       &     134427 & \multicolumn{2}{c}{--}       \\
        \MonCheriFlute               & 142069 & 2960 & $2\%$                  &    10705 & 56 & $0.5\%$                   &      135114 & 687 & $0.5\%$                  \\ \bottomrule
    \end{tabular}
\end{table*} 

\newcommand{\tworow}[1]{\multirow{-2}{*}{#1}}
\newcommand{\onerow}[1]{\multirow{1}{*}{#1}}

\begin{table*}[t!]
    \centering
    \caption{Performance cost on VCU118 @ 100MHz expressed as CoreMark test results. The CoreMark score for a processor is reported as CoreMark-iterations-per-second-per-core-MHz. The $\Delta$ is relative to \CheriFlute nocap results.}\label{tab:swcost}
    \resizebox{\textwidth}{!}{
    \begin{tabular}{r r rrc rrr rrr rrr c}\toprule
        & & \multicolumn{10}{c}{\textbf{CoreMark}} \\
        \cmidrule(lr){3-5}\cmidrule(lr){6-8}\cmidrule(lr){9-11}\cmidrule(lr){12-14}\cmidrule(lr){15-15}
        \rowcolor{white}                  &                                &    Binary size &       \multicolumn{2}{c}{$\Delta$} &         Total ticks & \multicolumn{2}{c}{$\Delta$} &      Total time (sec) &   \multicolumn{2}{c}{$\Delta$} & Iterations/sec &  \multicolumn{2}{c}{$\Delta$} &         Score \\ \midrule
        \rowcolor{white} \multicolumn{2}{c}{\CheriFlute}                 &                &                                    &                     &                                        &             &                                &                                                &               \\
        \rowcolor{Gray!10}                &               (baseline) nocap &          43728 &             \multicolumn{2}{c}{--} &          2704279286 &                 \multicolumn{2}{c}{--} &          27 &         \multicolumn{2}{c}{--} &          370 &          \multicolumn{2}{c}{--} &           3.7 \\
        \rowcolor{white}                  &                        purecap &          49872 &          6144 &          $14.05\%$ &          2878393823 &          174114537 &          $6.43\%$ &          28 &           1 &          $3.57\%$&          357 &          13 &          $3,51\%$ &          3.57 \\ \midrule
        \rowcolor{white} \multicolumn{2}{c}{\MonCheriFlute}              &                &                                    &                     &                    &                   &             &             &                  &              &             &                   &               \\
        \rowcolor{Gray!10}                &                          nocap &          43728 &             \multicolumn{2}{c}{--} &          2704301802 &              22516 &         $0.42\%$  &          27 &          0 &            $0\%$  &          370 &           0 &             $0\%$ &          3.7  \\
        \rowcolor{white}                  &                        purecap &          49872 &          6144 &          $14.05\%$ &          2878516285 &          174236999 &          $6.44\%$ &          28 &          1 &          $3.57\%$ &          357 &          13 &          $3.51\%$ &         3.57  \\
        \rowcolor{Gray!10} \onerow{\dOne} &     \WriteBeforeRead + purecap &          50224 &         6496 &          $14.85\%$ &           2949321554 &     245019752 &          $9.06\%$ &          29 &          2 &          $7.04\%$ &          344 &          26 &          $7.00\%$ &         3.44  \\
        \rowcolor{white}                  &     \WriteBeforeRead + purecap &                &                                    &                     &                    &                   &             &            &                   &              &             &                   &               \\
        \rowcolor{white}   \tworow{\dTwo} & excluding store linearization  & \tworow{49232} & \tworow{5504} & \tworow{$12.59\%$} & \tworow{2882957831} & \tworow{178678545} & \tworow{$6.61\%$} & \tworow{28} & \tworow{1} & \tworow{$3.57\%$} & \tworow{357} & \tworow{13} & \tworow{$3.51\%$} & \tworow{3.57} \\
\\ \bottomrule  
\end{tabular}}
\end{table*} 

We evaluated the \moncheri{} prototypes for functionality, security, performance, and area cost.
The functional and security evaluation (\Cref{sec:securityeval}) was performed on a \CC-enhanced QEMU-system-CHERI128 full-system emulator, \CheriFreeRtos~\cite{Almatary22}, which integrates \gls{cheri}-based compartmentalization, and a \WriteBeforeRead{} memory allocator (\Cref{sec:runtime}).
For performance and area cost, we extended the CHERI-RISC-V \gls{fpga} softcore~\cite{CTSRD24a} based on the open-source Bluespec~\cite{Nikhil04} RISC-V processor \gls{ip} with the \CP{} \gls{isa} extension (\Cref{sec:isa-ext}).
\ifnotabridged{}
This processor family includes Piccolo, Flute, and Tooba \gls{ip} cores.
Although sharing significant portions of \gls{bsv} code, our performance evaluation uses the 64-bit \gls{cheri}-RISC-V Flute (RV64ACDFIMSUxCHERI).
\fi
This prototype, \MonCheriFlute, was validated against RISC-V specifications using 229 RISC-V \gls{isa} tests~\cite{Newsome24}.

\subsection{Functional and Security Evaluation on QEMU}\label{sec:securityeval}

We used the U.S. \gls{nist} Juliet Test Suite~\cite{nsajuliet}, which includes thousands C/C++ of test cases that demonstrate common programming defects that lead to memory vulnerabilities, to evaluate \moncheri{}.
These tests are organized by CVE numbers, with ``bad'' versions exhibiting vulnerabilities and ``good'' versions showing patched code.
We focused on the CWE-457~\cite{MITRE06} (Use of Uninitialized Variable) C test cases, covering 560 bad and 560 corresponding good cases.
These examples cover a range of realistic scenarios, such as conditional control flows where variables might remain uninitialized in certain branches (e.g., as illustrated in \Cref{lst:example}), function calls that pass variables assumed to be initialized, and cases involving complex data types like arrays, pointers, and structures.

To extensively assess Mon CHERI's detection rate for uninitialized variable accesses, 
we made specific modifications to the Juliet test suite. 
Modern compilers like LLVM tend to optimize away uninitialized memory accesses 
or reject them outright due to their sophisticated static analysis capabilities. 
To prevent this behavior, we declared all variables as volatile, 
which stops the compiler from applying optimizations based on undefined behavior. 
This allows us to build the tests with optimization level \texttt{-O2} while ensuring all test bad cases trigger uninitialized memory accesses.

The tests were executed within \CheriFreeRtos{}, which uses an instrumented allocation \gls{api} to track memory accesses.
\CheriFreeRtos{} provides a ``compartmentalize and return'' mode that isolates each test case in a separate compartment, allowing test cases resulting in a \gls{cheri} protection fault to return control to the caller, which records the result and proceeds to the next test case.
The test cases were compiled with the \moncheri{}-enhanced CHERI-LLVM at optimization level \texttt{-O2} and with \WriteBeforeRead{} instrumentation and store linearization but without the \WriteBeforeRead{} optimization described in \Cref{sec:instrumentation}. 

The results in \Cref{tab:rwtable} show that \moncheri{} detects all ``bad'' cases (100\% true-positive rate) and reports only six false positives (1\% false-positive rate).
Upon closer inspection, these false positives stem from cases where uninitialized variables are copied, but never used.
Although these cases are technically valid violations of the \WriteBeforeRead{} policy, they do not pose a security risk, as data is immediately overwritten.
Analysis tools relying on taint propagation~\cite{Schwartz10} might not flag these cases, as the uninitialized memory does not propagate.
As \moncheri{} enforces policies at an architectural level, distinguishing between these benign cases and actual vulnerabilities is not currently possible.
All six false positives share this pattern, where the uninitialized memory is copied but later overwritten before being used.
We provide the source code for one of these cases in \Cref{app:juliet}, \Cref{lst:juliet-cwe457-63-good}.
Further, all six cases have straightforward software workarounds that can be applied, once detected, to initialize variables early with a default zero value to avoid uninitialized access.

We evaluated the effectiveness of the store linearization pass by comparing the detection rate for the Juliet tests instrumented with and without store linearization.
Without store linearization 119 out of 560 ``good'' test cases exhibit false positives (21\% false positive rate).
This demonstrates store linearization provides a significant improvement to the detection accuracy of \moncheri{}.
We also verified the \WriteBeforeRead{} optimization did not affect the detection accuracy as it only omits \CP{} instrumentation for variables that are statically verified to be fully initialized.

For comparison, we compiled the Juliet test suite with all relevant warnings enabled in GCC and Clang/LLVM.
GCC detected 170 out of 560 uninitialized cases (30\%) at \texttt{-O0} and 173 cases (31\%) at \texttt{-O2}.
Clang/LLVM detected 117 cases (21\%) regardless of optimization level.
Detection rates for the Valgrind and Dr. Memory dynamic analysis tools range from below 10\% to levels comparable to \moncheri, depending on the Juliet and compiler configurations.
Under the same configuration used for \moncheri, (volatile variables and Clang with \texttt{-O2}), Valgrind and Dr. Memory detected 400 cases (71\%) and 388 cases (69\%), respectively.

As the Juliet CWE457 tests do not exhibit patterns that would require \WriteBeforeExecute, \WriteBeforeReadOnly, \WriteBeforeExecuteOnly, or \WriteOnce{} \CPs{} we verified the functionality of these \CPs{} in the \CC-enhanced QEMU-system-CHERI128 implementation using purpose-built synthetic test cases.

\subsection{Performance and Area Evaluation on FPGA}

\paragraph{Area cost.}
We synthesized \MonCheriFlute{} at 100 MHz on an AMD Virtex UltraScale+ VCU-118 \gls{fpga}.
Compared to the \CheriFlute{} design, the area cost of \MonCheriFlute{} increased by only 2\%, a small cost considering the overhead of adding \gls{cheri}.
The majority of this additional logic is shared across many \CPs{}.

\paragraph{Performance cost.}
We integrated the \MonCheriFlute{} softcore into the BESSPIN-GFE security evaluation platform\cite{Podhradsky22}, which allows for a full-system evaluation of \moncheri{} performance.
\ifnotabridged{}
The GFE system includes the \MonCheriFlute{} softcore, a BootROM, Soft Reset and JTAG, UART, Ethernet/DMA, DDR4, and Flash controllers.
A host-based gdb debugger connects to the system over the USB/JTAG connector, and a host-based console connects over USB/UART.\@
\fi{}
We measured performance using the EEMBC CoreMark~\cite{EEMBC24} benchmark, running bare-metal on the GFE.\@
Although \MonCheriFlute{} supports all \CPs{} in \Cref{table:instructions} we focus in these experiments on \WriteBeforeRead{} as it is applicable to all variables in CoreMark.
Consequently, applying it to all variables in the CoreMark benchmark code provides a worst-case estimate of \moncheri{}'s performance impact.
As we expect other \CPs{} to only be applied to a subset of variables, their impact is a fraction of \WriteBeforeRead's.

\Cref{tab:swcost} compares the performance results \MonCheriFlute{} to the \CheriFlute{} across different configurations: no capability enforcement (no-cap), pure-capability mode (purecap), and \WriteBeforeRead{} enabled in pure-capability mode (\WriteBeforeRead{} + purecap \dOne).
The results show that the \WriteBeforeRead{} extension adds a modest \OverheadRelativeToCheri{} overhead over pure-capability mode, with minimal impact on baseline performance ($\approx$~0.4\%) when capability enforcement is disabled.
The combined performance impact of \WriteBeforeRead{} and \gls{cheri} pure-capability mode over the baseline performance with no capability enforcement is \OverheadRelativeTobaseline{}.

We also compared the performance of \MonCheriFlute{} with and without store linearization (\dTwo).
Although store linearization is necessary, as explained \Cref{sec:securityeval}, for the correctness of \WriteBeforeRead{} enforcement, disabling the hardware fault for this experiment allows us to compare the performance impact of the hardware changes with that of the store linearization program transformation.
Enabling store linearization resulted in most of the performance degradation observed in earlier tests ($\approx$~3.5\%), with the hardware changes contributing negligible additional overhead~($\approx$~0.2\%). 
This suggests that performance can be improved further by optimizing the store linearization strategy.

\paragraph{Microbenchmarks.}
To assess the impact of our store and load pipeline changes we microbenchmark stores and loads between \CheriFlute{} in purecap mode and \MonCheriFlute{} in \WriteBeforeExecute{} + purecap mode. 
The store microbenchmark writes a 256-element array, recording total ticks. 
The load microbenchmark, we measured read from an of equal size.
Here we report the difference in mean times for the experiment over 10 repetitions. 
We observed a negligible difference between \CheriFlute{} and \MonCheriFlute: $\approx5$ ticks for the load and $\approx30$ ticks for the store benchmark.

\glsunset{gm}

Finally, we evaluated the impact of \moncheri{} on the \gls{tlsf} allocator through microbenchmarks that allocate and free 1 MB of memory in chunks ranging from 32 bytes to 4 KiB.
The results, shown in Figure 5, indicate that the \WriteBeforeRead{}-enhanced \gls{tlsf} allocator introduces negligible performance overhead: $\approx$~0.1\% compared to purecap and $\approx$~1.5\% (\gls{gm}) compared to no-cap.
The overhead of \WriteBeforeRead{} + purecap is constant regardless of allocation size, while the overhead of zero-initialization increases linearly with allocation size.

\begin{figure}[t!]
    \includegraphics[width=\linewidth, scale=0.08]{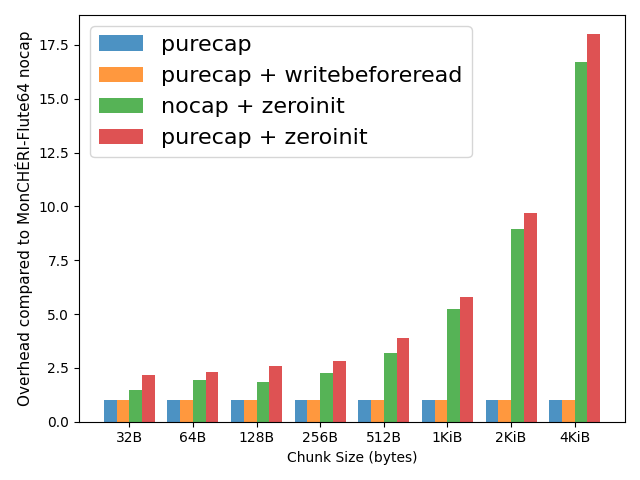}
    \caption{TLSF allocator microbenchmark. Bars show overhead relative to TLSF in \MonCheriFlute{} nocap mode.}\label{fig:microbenchmark}
\end{figure}

\section{Related Work}\label{sec:relatedwork}
\newcolumntype{R}[2]{
    >{\adjustbox{angle=#1,lap=\width-(#2)}\bgroup}
    l
    <{\egroup}
}
\newcommand*\rotate{\multicolumn{1}{R{30}{1em}}}
\newcommand*\column[1]{\multirow{2}{6em}{\centering\footnotesize #1}}
\newcommand*\leftalign{\rowcolor{Gray}\color{White}\footnotesize}
\newcommand*\rightalign{\quad\footnotesize}
\newcommand{\notemark}[1]{\textsuperscript{#1}}
\newcommand{\tablenote}[1]{\footnotesize{#1}}

\newcommand{\yes}{\cmark\xspace}
\newcommand{\no}{\xmark\xspace}
\newcommand{\maybe}{\textbf{?}\xspace}
\newcommand{\na}[1][\xspace]{--#1}

\begin{table*}
\centering
\caption{Related work}
\resizebox{\textwidth}{!}{
\begin{threeparttable}
\begin{tabular}{l c c c c c c}
    & \column{Support for\\ stack allocations}
    & \column{Support for\\ heap allocations}
    & \column{Usable with dynamic analysis}
    & \column{Unaffected by optimization level}
    & \column{Performance \\ overhead}
    & \column{Memory \\ overhead} \\
    & & & & & & \\
\toprule
    \leftalign{\textbf{Static code analysis}\notemark{1}}                                  &        &        &        &        &       &       \\
    \rightalign{\texttt{--Wuninitialized} (GCC~\cite{GCCdevelopercommunity14}}             & \yes{} & \yes{} & \yes{} & \no{}  & \na{} & \na{} \\
    \rightalign{\texttt{--Wmaybe-uninitialized} (GCC~\cite{GCCdevelopercommunity14})}      & \yes{} & \yes{} & \yes{} & \no{}  & \na{} & \na{} \\
    \rightalign{\texttt{--Wuninitialized} (Clang~\cite{LLVMteam24})}                       & \yes{} & \yes{} & \yes{} & \yes{}  & \na{} & \na{} \\
    \rightalign{\texttt{--Wsometimes-uninitialized} (Clang~\cite{LLVMteam24})}             & \yes{} & \yes{} & \yes{} & \yes{}  & \na{} & \na{} \\
    \midrule
    \leftalign{\textbf{Dynamic analysis}}                       &        &        &        &        &            &     \\
    \rightalign{Valgrind Memcheck~\cite{Seward05}}              & \yes{} & \yes{} & \yes{} & \no{} & 20$\times$ & yes \\
    \rightalign{Dr. Memory~\cite{Bruening11}}                   & \yes{} & \yes{} & \yes{} & \no{} & 10$\times$ & yes \\
    \midrule
    \leftalign{\textbf{Sanitizers}}                             &        &        &        &        &                      &     \\
    \rightalign{Memory Sanitizer~\cite{Stepanov15}}             & \yes{} & \yes{} & \yes{} & \yes{} & 2$\times$--4$\times$ & yes \\
    \midrule
    \leftalign{\textbf{Redundant execution}}                    &        &        &        &        &      &       \\
    \rightalign{DieHard~\cite{Berger06}}                        & \yes{} & \yes{} & \yes{} & \yes{} & $\approx$~40\% & yes   \\
    \rightalign{Differential Replay~\cite{Cao19}}               & \yes{} & \yes{} & \maybe{} & \yes{} & 22$\times$--24$\times$ & yes   \\
    \midrule
    \leftalign{\textbf{Automatic initialization}}               &               &          &        &        &                         &       \\
    \rightalign{Secure deallocation~\cite{Chow05}}              & \yes{}        & \yes{}   & \yes{} & \yes{} & <7\%                    & \na{} \\
    \rightalign{UniSan~\cite{Lu16}}                             & \yes{}        & \yes{}   & \no{}  & \yes{} & $\approx$~5\%           & \na{} \\
    \rightalign{SafeInit~\cite{Milburn17}}                      & \yes{}        & \yes{}   & \no{}  & \yes{} & $\approx$~5\%           & \na{} \\
    \rightalign{STACKLEAK~\cite{Popov18}}                       & *\notemark{2} & \no{}    & \no{}  & \yes{} & $\approx$~1\% -- 5\%    & \na{} \\
    \rightalign{initAll (\acrshort{msvc})\cite{Bialek20}}       & \yes{}        & \no{}    & \no{}  & \yes{} & $\approx$~10\%          & \na{} \\
    \rightalign{\TrivialAutoVarInit (GCC~\cite{GCCdevelopercommunity14}, Clang)}               & \yes{}        & \no{}    & \no{}  & \yes{} & $\approx$~1\%~\cite{Guelton23} -- 35\%~\cite{Nilsson23} & \na{} \\
    \midrule
    \leftalign{\textbf{Hardware-based detection}}               &               &          &        &        &               &       \\
    \rightalign{Uninitialized capabilities~\cite{Georges21}}    & *\notemark{4} & \no{}    & \yes{} & \yes{} & ?                                                                        & *\notemark{3} \\
    \rightalign{Capstone~\cite{Yu23}}                           & *\notemark{4} & \maybe{} & \yes{} & \yes{} & $\approx~50\%$\notemark{5}                                               & *\notemark{3} \\
    \rowcolor{Gray!10}\rightalign{\moncheri}                    & \yes{}        & \yes{}   & \yes{} & \yes{} & \OverheadRelativeToCheri $\ /\ $ \OverheadRelativeTobaseline\notemark{6} & *\notemark{3} \\
\bottomrule
\end{tabular}
\begin{tablenotes}
    \item[1] \tablenote{For conciseness, we include only compiler-based static analyzers focusing on uninitialized variable detection in \Cref{tab:relatedwork}.} \hfill $^2$ \tablenote{STACKLEAK protects the Linux kernel call stack after system calls.}
    \item[3] \tablenote{Memory overhead due to replacement of pointers with \gls{cheri} / Capstone capabilities.} \hfill $^5$ \tablenote{Overhead reported for the Capstone isolation model by Yu et al.~\cite{Yu23}.}
    \item[4] \tablenote{Uninitialized capabilities and Capstone protect stack frames at coarser granularity than what uninitialized variable detection for individual variables requires.}
    \item[6] \tablenote{Overhead for \moncheri given both excluding overhead for \gls{cheri} (purecap) and including overhead for \gls{cheri}, based on \Cref{tab:swcost}.}
\end{tablenotes}
\end{threeparttable}}
\label{tab:relatedwork}
\end{table*}

Various techniques for detecting the use of uninitialized variables are routinely used in modern software development.
We categorize the existing approaches into six categories, as shown in \Cref{tab:relatedwork}.
In this section, we compare \CCs{} and \moncheri{} to existing approaches.

\paragraph{Static analysis.}
Static analysis evaluates a program's code without executing it. 
By analyzing the code's structure and syntax, compilers, and dedicated static analysis tools can detect potential errors, security issues, and coding standard violations.
Static analysis is performed early in development, allowing developers to address problems proactively. 

Most major C/C++ compilers, including GCC~\cite{GCCdevelopercommunity14}, Clang\cite{LLVMteam24}, Intel \gls{dpc}, and \gls{msvc} support compile-time checks for uninitialized variables.
Static analysis tools  such as Adlint~\cite{Yutaka15}, Clang-Check~\cite{LLVMteam24b}, Clang-Tidy~\cite{LLVMTeam24a}, CodeSonar~\cite{CodeSecure23}, Coverity Scan~\cite{Synopsys23}, CppCheck~\cite{Marjamaki24}, Flawfinder~\cite{Wheeler07}, Frama-C~\cite{Kirchner15}, IKOS~\cite{Brat14}, Infer~\cite{Facebook16}, and LCLint/Splint~\cite{Evans02} can also check for uninitialized variables in C/C++ code.
\ifnotabridged{}
These tools, particularly commercial tools focused on secure software development, are commonly referred to and marketed as \gls{sast}.
\fi

However, all static analysis is limited by Rice’s Theorem~\cite{Rice53}, which states that analyzing non-trivial properties of program behavior is undecidable for Turing-complete languages.
\ifnotabridged{}
This implies that static detection of uninitialized variables is equivalent to solving the halting problem.
\fi
As a result, static methods are approximations, balancing verbosity with false positives and analysis time.

\paragraph{Dynamic analysis.}
Dynamic analysis examines program behavior during execution.
Unlike static analysis, which analyzes code without running it, dynamic analysis monitors run-time characteristics such as performance, memory usage, and interaction with system resources.
It can, therefore, identify bugs, security vulnerabilities, and performance bottlenecks that may not be evident through static analysis.
Tools like Valgrind Memcheck~\cite{Seward05} and Dr.~Memory~\cite{Bruening11} use dynamic instrumentation frameworks (e.g, DynamoRIO~\cite{Bruening04}) to track memory accesses and detect use of uninitialized variables.
\ifnotabridged{}
These frameworks act as process virtual machines, interposing and transforming original program instructions before they get executed by the hardware.
This enables dynamic analysis to freely transform the target program and add extra instrumentation around program instructions to keep track of memory accesses
\fi

Dynamic analysis, however, incurs significant performance overhead, often degrading program speed by 10$\times$ to 20$\times$, making it impractical for continuous use.
These tools also struggle to accurately identify the origin of uninitialized memory.
For example, Memcheck traces uninitialized variables to heap blocks or stack allocations that occur in a particular function, but may not always pinpoint the exact source.
Dr.~Memory, meanwhile, does not detect uninitialized variables smaller than a machine word.

\paragraph{Sanitizers.}
Sanitizers are compiler-based tools designed to detect memory-safety, concurrency, and undefined behavior issues in C and C++ programs.
They intercept memory accesses via compile-time instrumentation, offering higher efficiency and accuracy than dynamic analysis tools.
Sanitizers like MemorySanitizer\cite{Stepanov15} detect uninitialized
stack- and heap-allocated memory at individual bit granularity, with less
overhead  (2$\times$ to 4$\times$ slowdown) compared to Valgrind's one-order-of-magnitude slower dynamic analysis~\cite{Nethercote07}. 

However, like Valgrind Memcheck, MemorySanitizer only reports uninitialized values that affect control flow, which limits its effectiveness in identifying memory issues related to information disclosure to, e.g., uninitialized data that overlaps with previously allocated pointers and which can reveal information to bypass \gls{aslr}.

\paragraph{Redundant execution techniques.}
Redundant execution techniques, such as DieHard~\cite{Berger06}, enhance memory safety by using randomized memory allocation and replication.
DieHard scatters memory allocations across a large heap, reducing the chances of uninitialized memory being adjacent to other active regions. 
By comparing execution across multiple program replicas, DieHard can detect discrepancies caused by uninitialized memory reads.
Similarly, differential replay~\cite{Cao19} captures execution traces and replays them with varied initial memory states.
\ifnotabridged{}
These techniques assume that correctly initialized variables will produce consistent outputs across runs, while uninitialized variables will lead to variations due to differences in their initial memory state.
By identifying discrepancies between execution instances, DieHard and differential replay can pinpoint instances where uninitialized variables are affecting the program's behavior.
\fi

Multi-variant execution~\cite{Cox06, Salamat09, Jackson10, Koning16, Volckaert16, Coppens18} generalizes this concept, running multiple functional equivalent, but independently developed, programs in parallel.
This principle has been applied to safety-critical software in various
domains, such as train switching and flight control systems, electronic
voting, and specialized software testing, such as detecting zero-day
exploits and kernel information leaks~\cite{Osterlund19}.
However, replicating compute instances and I/O across each variant is resource-intensive and impractical for general-purpose use.

\paragraph{Automatic initialization.}
Automatic initialization approaches, such as UniSan~\cite{Lu16} and SafeInit\cite{Milburn17}, automatically set variables to default values, mitigating uninitialized memory issues.
These methods introduce performance overhead, particularly with large allocations~\cite{Nilsson23}, and may miss issues with non-stack variables.
For example, Microsoft MSVC's initAll~\cite{Bialek20} and the \TrivialAutoVarInit{} option in GCC and Clang focus on stack variables, but their their use is limited by their performance penalties~\cite{Bialek20,Guelton23}.

Automatic initialization can also interfere with dynamic analysis tools and sanitizers, masking issues with uninitialized variables that could be detected and fixed, making it less suitable for debugging and software testing~\cite{OpenSSFcontributors24}.
Chow et al.~\cite{Chow05} propose a secure deallocation technique to zero memory upon function exit, though it shares similar drawbacks, introducing overhead at the end of object lifetimes.
The Linux Kernel implements a similar scheme, STACKLEAK~\cite{Popov18}, that clears the kernel stack at the end of system calls.
This mitigates the impact of information leakage bugs, although it can impact system performance by up to 5\%.

\paragraph{Hardware-based detection.}
Hardware-based detection methods, such as Georges et al.'s uninitialized capabilities~\cite{Georges21} and Capstone~\cite{Yu23}, attempt to address uninitialized memory issues by simply introducing a new permission to \gls{cheri}.
\ifnotabridged{}
Unlike \moncheri, which uses an operation-specific bound to track written portions of the address space, uninitialized capabilities grant read permissions in the range $\left[\CCAddress, \gls{top}\right]$ and write permission in $\left[\gls{base},\gls{top}\right]$ where $\CCAddress$ is the baseline address, \gls{base} is the capability bound base, and \gls{top} is the capability bound top.
\fi
However, this approach has significant limitations.

Firstly, the use of uninitialized capabilities requires software to derive new capability\ifnotabridged{}, with an adjusted $\CCAddress$,\fi{} with each write operation, introducing complexity in managing memory permissions. 
This limits the utility of uninitialized capabilities to a secure calling convention that enhances local stack frame encapsulation, first proposed by Skorstengaard et al.~\cite{Skorstengaard19}, by additionally protecting against uninitialized stack reads.

Secondly, uninitialized capabilities only support \WriteBeforeRead{} semantics, meaning they do not provide protection against other forms of uninitialized memory access.
This limited expressibility reduces the utility of the approach, particularly when compared to more flexible solutions like \moncheri{}, which supports a broader range of access control policies as well as protection for both stack, heap, and other types of memory allocations.

Capstone~\cite{Yu23} is a redesign of the \gls{cheri} capability model that enables broader memory isolation and attempts to generalize uninitialized capabilities. 
It addresses the first drawback of Georges et al.’s method by introducing self-incrementing write semantics.
However, like Georges et al.’s uninitialized capabilities, Capstone shares the limitation of only addressing \WriteBeforeRead{} scenarios.
Capstone also suffers from a major drawback of its own: it incurs a significant performance overhead of up to 50\%.

Additionally, in Capstone, fully initialized capabilities must be explicitly promoted to regular capabilities.
Either the developer or compiler must understand when a capability is expected to be fully initialized, in order to promote it.
In contrast, due to the way \CC{} semantics are designed (\Cref{sec:cc}), a fully initialized \CC{} is equivalent to the corresponding \gls{cheri} capability.

These prior hardware-based approaches also face integration challenges.
Uninitialized capabilities have only been simulated on the obsolete CHERI-MIPS \gls{isa}~\cite{Huyghebaert20}.
Capstone, on the other hand, comes with invasive changes to the established \gls{cheri} architecture and high overhead, making it impractical for real-world applications, especially in performance-critical systems.

\paragraph{Comparison with \moncheri.}
Our evaluation demonstrates that \CPs{}, particularly \WriteBeforeRead{} capabilities in \moncheri{}, offer high detection accuracy with minimal false positives when used in isolation.
While static analysis is valuable for early detection, tools like dynamic analysis, sanitizers, and \CPs{} require runtime errors to be exercised.
We believe that \WriteBeforeRead{} \CPs{} can complement static analysis by improving the detection of uninitialized memory issues.
In \Cref{sec:discussion}, we discuss how static analysis can be used with \CP{} instrumentation to further optimize \CP{} performance.

Our assessment of \moncheri{} is based on extensive evaluation using standard public sector and industry benchmarks on prototypes based on the QEMU full-system emulator and the \MonCheriFlute{} softcore (\Cref{sec:evaluation}).
To our knowledge, no practical implementation nor compiler support for Georges et al.’s uninitialized capabilities is available to enable a fair comparison with \moncheri{} on \gls{fpga}.
We argue that \CCs{}, carefully designed to impose only minimal changes to the \gls{cheri} capability representation (\Cref{sec:impl-cc}), have a better chance of real-world adoption than more invasive proposals, such as Capstone.

Hardware-enforced \CPs{} can improve detection performance for uninitialized memory issues similar to how memory tagging~\cite{Arm19} has enhanced memory-safety sanitizers~\cite{Serebryany18}.
Moreover, \CPs{} offer an alternative to automatic initialization (see \Cref{sec:discussion}), emulating it without the associated drawbacks.
Lastly, \CPs{} enable novel memory access control policies, such as \WriteBeforeExecuteOnly{}, which provide similar benefits to memory with special-purpose features.

\section{Discussion and Future Work}\label{sec:discussion}\label{sec:futurework}
Here, we discuss limitations of the current \moncheri{} prototype, alternate designs, and suggest future research.

\paragraph{Leveraging compiler-based static analysis.}
In \Cref{sec:instrumentation} we showed how \WriteBeforeRead{} \CPs{} can be omitted for variables that are statically verified to be initialized.
Existing compiler heuristics could be used to further optimize \WriteBeforeRead{}, e.g., by omitting \WriteBeforeRead{} \CPs{} when a variable is determined initialized by GCC's \texttt{-Wuninitialized}, or instrument only variables identified as potentially uninitialized in certain code paths by GCC's \texttt{-Wmaybe-uninitialized}.
We leave further optimizations as future work as Clang/LLVM, which our \CC{}-enhanced compiler prototype is based on, does not currently replicate GCC's \texttt{-Wmaybe-uninitialized} heuristic. 
Clang's alternative \texttt{-Wsometimes-uninitialized} is more conservative as it only issues warnings when the conditions under which a variable is left uninitialized are known.
Developers can already correct such cases based on the emitted warnings.
We believe the \WriteBeforeRead{} \CPs{} are useful in complementing compiler-based analysis, covering cases where the analysis is inconclusive.

\paragraph{Inter-function store linearization.}
The store linearization pass (\Cref{sec:instrumentation}) is currently limited to intra-function analysis, such as when \WriteBeforeRead{} \CCs{} are used within functions or passed to callees.
However, currently propagating \CCs{} from a callee back to the caller must be done explicitly to avoid the conditions explained in \Cref{sec:instrumentation} to occur across function boundaries.
Solving inter-function store linearization is simpler in languages that enforce \emph{borrowing}, where only one mutable reference to an allocation can exist at a time.
Borrowing is prominent in Rust~\cite{Klabnik18}, but similar, compiler-enforced borrow checking has been proposed for C~\cite{Silva24} and C++\cite{Sutter19,Lado24,Baxter24} as well.
Future work should explore borrow-checker-aided full program \CC{} linearization.

\paragraph{Emulating automatic initialization.}
In \Cref{sec:relatedwork}, we suggest \WriteBeforeRead{} \CPs{} could emulate automatic initialization, avoiding its drawbacks.
Instead of issuing a hardware protection fault, a load to an address outside the \CC's \gls{opbound} could set the destination register to zero (or a default pattern).
This mimics automatic initialization without the overhead of pre-initializing memory.

Another drawback of automatic initialization, discussed in \Cref{sec:relatedwork}, is interference with dynamic analysis and sanitizers.
Initialization emulation could be controlled via a hardware configuration, allowing it to be toggled on or off.
This allows the same \CP-instrumented program binary to be used in production and for testing, depending on the configuration.

\paragraph{Overlapping capability permission bits.}
In \Cref{sec:isa-ext} we explain how \moncheri{} reuses software-defined
permission bits in the \gls{cheri} capability format.
A drawback of the enumerator-based \CCOpPerms{} representation is that it makes \CPs{}, and possible software-defined permissions, mutually exclusive.

To avoid this drawback, we considered an alternative based on that \CPs{} always describe a subset of the conventional \gls{cheri} permissions, e.g., \WriteBeforeRead{} confers the same access as \Read{} when \gls{optop} $=$ \gls{top}.
In this alternative, a \emph{conditional control bit}, \CCConditional{} is assigned from the reserved, but unused, capability bits to indicate whether the capability is in \emph{conditional} mode.
In conditional mode, i.e., when $\CCConditional = 1$, the \CCHWPerms{} bits \Read{}, \Write{}, and \Execute{} represent the corresponding \CPs: \WriteBeforeRead{}, \WriteOnce{}, and \WriteBeforeRead{} respectively.
This allows \CP{} bits to overlay conventional \gls{cheri} permissions bits without losing expressivity.
The \WriteBeforeReadOnly, and \WriteBeforeExecuteOnly{} \CPs{} could be overlayed similarly with an additional \emph{exclusive permission} bit.

\paragraph{Non-sequentially written memory.}
In \Cref{sec:isa-ext}, we assume that memory accessible via \CPs{} is written sequentially.
This holds for most data types and operations like \texttt{memcpy()} and \texttt{memset()}, but not for C structures that are initialized field-by-field or contain uninitialized padding.
Software workarounds like \texttt{\#pragma pack} with ordered field-initialization or an initial \texttt{memset()} before individual fields are set can prevent false positives from field-by-field access.

Alternatively, \CCs{} can treat the \gls{optop} value as a bitmap, where each bit describes the initialization state of a memory segment.
Memory can be segmented into equal chunks or structured data fields.
If the 16-bit \gls{optop} field is insufficient, it could point to a larger bitmap in shadow memory, similar to how MemorySanitizer\cite{Stepanov15} tracks initialized memory.
However, we believe the existence of straightforward software workarounds makes the overhead of complex tracking solutions unnecessary and thus leave exploring solutions for tracking non-sequentially initialized memory outside the scope of this work.

\glsreset{base}
\glsreset{top}
\glsreset{optop}

\paragraph{Supporting $\CCInternalExponent > 2$.}
In \Cref{sec:impl-cc}, we explain that the current operation bounds encoding restricts \CCs to $\CCInternalExponent \le 2$.
Similar to how the \gls{cheri} Concentrate encoding sacrifices \emph{alignment} precision as allocation sizes increase, the \gls{optop} can sacrifice \emph{write} precision as $\CCInternalExponent$ increases.
At an architectural level, this is achieved by zero-padding the least significant bits of $\CCExtOp$, akin to handling \CCBase{} and \CCTop.
When the precision of the least significant stored bit in $\CCExtOp$ exceeds the ability to express writes to individual bytes, halfwords, or words, corresponding store instructions (\texttt{s[bhw]}) are disabled for that \CC.
Extending writes beyond doubleword precision requires a variant of \texttt{SetOpBounds}, which extends \gls{optop} under the condition that the two preceding instructions have been writes reaching a target granularity.

\section{Conclusion}\label{sec:conclusion}
\glsresetall{}
\glsunset{cheri}

This paper presents \moncheri{}, a novel extension to the \gls{cheri} architecture that addresses uninitialized memory errors that account for $\approx$~10\% of all memory vulnerabilities.

By introducing \CCs{}, \moncheri{} enables precise run-time detection of uninitialized memory access at instruction-level, with minimal performance overhead.
Our extensive evaluation on the \moncheri{} QEMU-system-CHERI128 emulator
and FPGA-based \MonCheriFlute{} prototype shows that \moncheri{} achieves a
100\% true-positive rate while maintaining a low, 1\%, false-positive rate
on the Juliet test suite, and incurs only a \OverheadRelativeToCheri{} performance overhead for the \WriteBeforeRead{} extension.
 
Our comparison with state-of-the-art solutions, including static analysis tools, sanitizers, and other hardware-based detection techniques demonstrates that \moncheri{} complements static analysis and provides additional coverage where static analysis alone falls short.
Moreover, \CPs{} enable novel memory access control policies, while being carefully designed to impose only minimal changes to the \gls{cheri} capability representation and thus have a better chance of real-world adoption than previous proposals that provide only \WriteBeforeRead{} semantics, with significant limitations or invasive changes. 

Looking ahead, we plan to extend \moncheri{} capabilities by integrating it with broader real-world use cases, emulation of automatic initialization, explore further optimization strategies, and reducing false positives in edge cases though compiler improvements.

\ifnotanonymous
\section*{Acknowledgments}
Firstly, we would like to extend our gratitude to Prof. N. Asokan and Hossam ElAtali at the University of Waterloo for allowing us to conduct part of our evaluation on the group's \acrshort{fpga} equipment.

We are equally grateful for the guidance provided by the University of Cambridge Computer Laboratory Security Group members, particularly Dr. Hesham Almatary, Dr. Jonathan Woodruff, Jessica Clarke, and Prof. Robert Watson.
We also wish to thank our colleagues at Ericsson: Christoph Baumann, Patrik Ekdahl, Peter Svensson, and Santeri Paavolainen for supporting various aspects of our research.
Lastly, we appreciate the valuable discussions we've had with regards to this work with Hans Liljestrand, Adriaan Jacobs, and Fatih Aşağıdağ.

This work has received funding under EU H2020 MSCA-ITN action 5GhOSTS, grant agreement no. 814035, by the Research Fund KU Leuven, by the Flemish Research Programme Cybersecurity, and by the CyberExcellence programme of the Walloon Region, Belgium.

\fi
\enlargethispage{1em}
\bibliographystyle{IEEEtran}
\bibliography{main, local}

\begin{thebibliography}{10}
\providecommand{\url}[1]{#1}
\csname url@samestyle\endcsname
\providecommand{\newblock}{\relax}
\providecommand{\bibinfo}[2]{#2}
\providecommand{\BIBentrySTDinterwordspacing}{\spaceskip=0pt\relax}
\providecommand{\BIBentryALTinterwordstretchfactor}{4}
\providecommand{\BIBentryALTinterwordspacing}{\spaceskip=\fontdimen2\font plus
\BIBentryALTinterwordstretchfactor\fontdimen3\font minus
  \fontdimen4\font\relax}
\providecommand{\BIBforeignlanguage}[2]{{%
\expandafter\ifx\csname l@#1\endcsname\relax
\typeout{** WARNING: IEEEtran.bst: No hyphenation pattern has been}%
\typeout{** loaded for the language `#1'. Using the pattern for}%
\typeout{** the default language instead.}%
\else
\language=\csname l@#1\endcsname
\fi
#2}}
\providecommand{\BIBdecl}{\relax}
\BIBdecl

\bibitem{Cho20}
H.~Cho \emph{et~al.}, ``Exploiting {{Uses}} of {{Uninitialized Stack
  Variables}} in {{Linux Kernels}} to {{Leak Kernel Pointers}},'' in \emph{14th
  {{USENIX Workshop}} on {{Offensive Technologies}} ({{WOOT}} 20)}, ser.
  {{WOOT}} '20.\hskip 1em plus 0.5em minus 0.4em\relax USENIX Association, Aug.
  2020.

\bibitem{Sullivan14a}
\BIBentryALTinterwordspacing
N.~Sullivan, ``The {{Results}} of the {{CloudFlare Challenge}},'' Nov. 2014,
  (accessed 2024-07-05). [Online]. Available:
  \url{https://blog.cloudflare.com/the-results-of-the-cloudflare-challenge}
\BIBentrySTDinterwordspacing

\bibitem{Lu17}
\BIBentryALTinterwordspacing
K.~Lu \emph{et~al.}, ``Unleashing {{Use-Before-Initialization Vulnerabilities}}
  in the {{Linux Kernel Using Targeted Stack Spraying}},'' in \emph{Proceedings
  2017 {{Network}} and {{Distributed System Security Symposium}}}.\hskip 1em
  plus 0.5em minus 0.4em\relax San Diego, CA: Internet Society, 2017. [Online].
  Available:
  \url{https://www.ndss-symposium.org/ndss2017/ndss-2017-programme/unleashing-use-initialization-vulnerabilities-linux-kernel-using-targeted-stack-spraying/}
\BIBentrySTDinterwordspacing

\bibitem{Joly20}
\BIBentryALTinterwordspacing
N.~Joly, S.~ElSherei, and S.~Amar, ``Security {{Analysis}} of {{CHERI ISA}},''
  Microsoft Security Response Center, Tech. Rep., Oct. 2020. [Online].
  Available:
  \url{https://msrc.microsoft.com/blog/2020/10/security-analysis-of-cheri-isa/}
\BIBentrySTDinterwordspacing

\bibitem{Bialek20}
\BIBentryALTinterwordspacing
J.~Bialek, ``Solving {{Uninitialized Stack Memory}} on {{Windows}} {\textbar}
  {{MSRC Blog}} {\textbar} {{Microsoft Security Response Center}},'' Mar. 2020,
  (accessed 2024-06-27). [Online]. Available:
  \url{https://msrc.microsoft.com/blog/2020/05/solving-uninitialized-stack-memory-on-windows/}
\BIBentrySTDinterwordspacing

\bibitem{Sutter24}
\BIBentryALTinterwordspacing
H.~Sutter, ``C++ safety, in context,'' Mar. 2024, (accessed 2024-03-14).
  [Online]. Available:
  \url{https://herbsutter.com/2024/03/11/safety-in-context/}
\BIBentrySTDinterwordspacing

\bibitem{Zhao24}
\BIBentryALTinterwordspacing
L.~Zhao \emph{et~al.}, ``A {{Survey}} of {{Hardware Improvements}} to {{Secure
  Program Execution}},'' \emph{ACM Comput. Surv.}, p.~35, Jun. 2024. [Online].
  Available: \url{https://doi.org/10.1145/3672392}
\BIBentrySTDinterwordspacing

\bibitem{Qualcomm17}
\BIBentryALTinterwordspacing
Qualcomm, ``Pointer {{Authentication}} on {{ARMv8}}.3: {{Design}} and
  {{Analysis}} of the {{New Software Security Instructions}},'' Whitepaper,
  Jan. 2017. [Online]. Available:
  \url{https://www.qualcomm.com/content/dam/qcomm-martech/dm-assets/documents/pointer-auth-v7.pdf}
\BIBentrySTDinterwordspacing

\bibitem{Arm19}
\BIBentryALTinterwordspacing
Arm, ``Armv8.5-{{A Memory Tagging Extension}},'' Whitepaper, Aug. 2019.
  [Online]. Available:
  \url{https://developer.arm.com/-/media/Arm%20Developer%20Community/PDF/Arm_Memory_Tagging_Extension_Whitepaper.pdf}
\BIBentrySTDinterwordspacing

\bibitem{Intel20}
\BIBentryALTinterwordspacing
Intel, ``A {{Technical Look}} at {{Intel}}'s {{Control-flow Enforcement
  Technology}},'' Jun. 2020, (accessed 2023-10-28). [Online]. Available:
  \url{https://www.intel.com/content/www/us/en/developer/articles/technical/technical-look-control-flow-enforcement-technology.html}
\BIBentrySTDinterwordspacing

\bibitem{Woodruff14}
\BIBentryALTinterwordspacing
J.~Woodruff \emph{et~al.}, ``The {{CHERI}} capability model: {{Revisiting
  RISC}} in an age of risk,'' in \emph{2014 {{ACM}}/{{IEEE}} 41st
  {{International Symposium}} on {{Computer Architecture}} ({{ISCA}})}, Jun.
  2014, pp. 457--468. [Online]. Available:
  \url{https://ieeexplore.ieee.org/document/6853201}
\BIBentrySTDinterwordspacing

\bibitem{WesleyFilardo20}
\BIBentryALTinterwordspacing
N.~Wesley~Filardo \emph{et~al.}, ``Cornucopia: {{Temporal Safety}} for {{CHERI
  Heaps}},'' in \emph{2020 {{IEEE Symposium}} on {{Security}} and {{Privacy}}
  ({{SP}})}, May 2020, pp. 608--625. [Online]. Available:
  \url{https://ieeexplore.ieee.org/document/9152640}
\BIBentrySTDinterwordspacing

\bibitem{Watson19}
\BIBentryALTinterwordspacing
R.~N.~M. Watson \emph{et~al.}, ``An {{Introduction}} to {{CHERI}},'' Computer
  Laboratory, University of Cambridge, Technical {{Report}} UCAM-CL-TR-941,
  Sep. 2019. [Online]. Available:
  \url{https://www.cl.cam.ac.uk/techreports/UCAM-CL-TR-941.pdf}
\BIBentrySTDinterwordspacing

\bibitem{Levy84}
\BIBentryALTinterwordspacing
H.~M. Levy, \emph{Capability-{{Based Computer Systems}}}.\hskip 1em plus 0.5em
  minus 0.4em\relax USA: Butterworth-Heinemann, 1984. [Online]. Available:
  \url{https://homes.cs.washington.edu/~levy/capabook/}
\BIBentrySTDinterwordspacing

\bibitem{Dennis66}
\BIBentryALTinterwordspacing
J.~B. Dennis and E.~C. Van~Horn, ``Programming semantics for multiprogrammed
  computations,'' \emph{Communications of the ACM}, vol.~9, no.~3, pp.
  143--155, Mar. 1966. [Online]. Available:
  \url{https://dl.acm.org/doi/10.1145/365230.365252}
\BIBentrySTDinterwordspacing

\bibitem{Watson23}
\BIBentryALTinterwordspacing
R.~N.~M. Watson \emph{et~al.}, ``Capability {{Hardware Enhanced RISC
  Instructions}}: {{CHERI Instruction-Set Architecture}} ({{Version}} 9),''
  Computer Laboratory, University of Cambridge, Technical {{Report}}
  UCAM-CL-TR-987, Sep. 2023. [Online]. Available:
  \url{https://www.cl.cam.ac.uk/techreports/UCAM-CL-TR-987.html}
\BIBentrySTDinterwordspacing

\bibitem{Grisenthwaite22}
\BIBentryALTinterwordspacing
R.~Grisenthwaite, ``Arm {{Morello Evaluation Platform}} -{{Validating
  CHERI-based Security}} in a {{High-performance System}},'' in \emph{2022
  {{IEEE Hot Chips}} 34 {{Symposium}} ({{HCS}})}, Aug. 2022, pp. 1--22.
  [Online]. Available: \url{https://ieeexplore.ieee.org/document/9895591}
\BIBentrySTDinterwordspacing

\bibitem{Amar23}
\BIBentryALTinterwordspacing
S.~Amar \emph{et~al.}, ``{{CHERIoT}}: {{Complete Memory Safety}} for {{Embedded
  Devices}},'' in \emph{56th {{Annual IEEE}}/{{ACM International Symposium}} on
  {{Microarchitecture}}}.\hskip 1em plus 0.5em minus 0.4em\relax Toronto ON
  Canada: ACM, Oct. 2023, pp. 641--653. [Online]. Available:
  \url{https://dl.acm.org/doi/10.1145/3613424.3614266}
\BIBentrySTDinterwordspacing

\bibitem{Woodruff19}
\BIBentryALTinterwordspacing
J.~Woodruff \emph{et~al.}, ``{{CHERI Concentrate}}: {{Practical Compressed
  Capabilities}},'' \emph{IEEE Transactions on Computers}, vol.~68, no.~10, pp.
  1455--1469, Oct. 2019. [Online]. Available:
  \url{https://ieeexplore.ieee.org/document/8703061}
\BIBentrySTDinterwordspacing

\bibitem{Xia19}
\BIBentryALTinterwordspacing
H.~Xia \emph{et~al.}, ``{{CHERIvoke}}: {{Characterising Pointer Revocation}}
  using {{CHERI Capabilities}} for {{Temporal Memory Safety}},'' in
  \emph{Proceedings of the 52nd {{Annual IEEE}}/{{ACM International Symposium}}
  on {{Microarchitecture}}}, ser. {{MICRO}} '52.\hskip 1em plus 0.5em minus
  0.4em\relax New York, NY, USA: Association for Computing Machinery, Oct.
  2019, pp. 545--557. [Online]. Available:
  \url{https://dl.acm.org/doi/10.1145/3352460.3358288}
\BIBentrySTDinterwordspacing

\bibitem{Filardo24}
\BIBentryALTinterwordspacing
N.~W. Filardo \emph{et~al.}, ``Cornucopia {{Reloaded}}: {{Load Barriers}} for
  {{CHERI Heap Temporal Safety}},'' in \emph{Proceedings of the 29th {{ACM
  International Conference}} on {{Architectural Support}} for {{Programming
  Languages}} and {{Operating Systems}}, {{Volume}} 2}.\hskip 1em plus 0.5em
  minus 0.4em\relax La Jolla CA USA: ACM, Apr. 2024, pp. 251--268. [Online].
  Available: \url{https://dl.acm.org/doi/10.1145/3620665.3640416}
\BIBentrySTDinterwordspacing

\bibitem{Chisnall17}
\BIBentryALTinterwordspacing
D.~Chisnall \emph{et~al.}, ``{{CHERI JNI}}: {{Sinking}} the {{Java Security
  Model}} into the {{C}},'' in \emph{Proceedings of the {{Twenty-Second
  International Conference}} on {{Architectural Support}} for {{Programming
  Languages}} and {{Operating Systems}}}, ser. {{ASPLOS}} '17.\hskip 1em plus
  0.5em minus 0.4em\relax New York, NY, USA: Association for Computing
  Machinery, Apr. 2017, pp. 569--583. [Online]. Available:
  \url{https://dl.acm.org/doi/10.1145/3037697.3037725}
\BIBentrySTDinterwordspacing

\bibitem{Milburn17}
\BIBentryALTinterwordspacing
A.~Milburn, H.~Bos, and C.~Giuffrida, ``{{SafeInit}}: {{Comprehensive}} and
  {{Practical Mitigation}} of {{Uninitialized Read Vulnerabilities}},'' in
  \emph{Proceedings 2017 {{Network}} and {{Distributed System Security
  Symposium}}}.\hskip 1em plus 0.5em minus 0.4em\relax San Diego, CA: Internet
  Society, 2017. [Online]. Available:
  \url{https://www.ndss-symposium.org/ndss2017/ndss-2017-programme/safelnit-comprehensive-and-practical-mitigation-uninitialized-read-vulnerabilities/}
\BIBentrySTDinterwordspacing

\bibitem{King15}
\BIBentryALTinterwordspacing
C.~I. King, ``Crypto: Mv\_cesa - ensure backlog is initialised {$\cdot$}
  torvalds/linux@1a92b2b {$\cdot$} {{GitHub}},'' Apr. 2015. [Online].
  Available:
  \url{https://github.com/torvalds/linux/commit/1a92b2ba339221a4afee43adf125fcc9a41353f7}
\BIBentrySTDinterwordspacing

\bibitem{Georges21}
\BIBentryALTinterwordspacing
A.~L. Georges \emph{et~al.}, ``Efficient and provable local capability
  revocation using uninitialized capabilities,'' \emph{Proceedings of the ACM
  on Programming Languages}, vol.~5, no. POPL, pp. 6:1--6:30, Jan. 2021.
  [Online]. Available: \url{https://dl.acm.org/doi/10.1145/3434287}
\BIBentrySTDinterwordspacing

\bibitem{Huyghebaert20}
\BIBentryALTinterwordspacing
S.~Huyghebaert \emph{et~al.}, ``Uninitialized {{Capabilities}},'' Jun. 2020.
  [Online]. Available: \url{http://arxiv.org/abs/2006.01608}
\BIBentrySTDinterwordspacing

\bibitem{Yu23}
\BIBentryALTinterwordspacing
J.~Z. Yu \emph{et~al.}, ``Capstone: {{A Capability-based Foundation}} for
  {{Trustless Secure Memory Access}},'' in \emph{32nd {{USENIX Security
  Symposium}} ({{USENIX Security}} 23)}, 2023, pp. 787--804. [Online].
  Available:
  \url{https://www.usenix.org/conference/usenixsecurity23/presentation/yu-jason}
\BIBentrySTDinterwordspacing

\bibitem{Shacham07}
\BIBentryALTinterwordspacing
H.~Shacham, ``The geometry of innocent flesh on the bone: Return-into-libc
  without function calls (on the x86),'' in \emph{Proceedings of the 14th
  {{ACM}} Conference on {{Computer}} and Communications Security}, ser. {{CCS}}
  '07.\hskip 1em plus 0.5em minus 0.4em\relax New York, NY, USA: Association
  for Computing Machinery, Oct. 2007, pp. 552--561. [Online]. Available:
  \url{https://doi.org/10.1145/1315245.1315313}
\BIBentrySTDinterwordspacing

\bibitem{Denis-Courmont20}
\BIBentryALTinterwordspacing
R.~{Denis-Courmont} \emph{et~al.}, ``Camouflage: {{Hardware-assisted CFI}} for
  the {{ARM Linux}} kernel,'' in \emph{2020 57th {{ACM}}/{{IEEE Design
  Automation Conference}} ({{DAC}})}, Jul. 2020, pp. 1--6. [Online]. Available:
  \url{https://ieeexplore.ieee.org/document/9218535}
\BIBentrySTDinterwordspacing

\bibitem{CTSRD24}
\BIBentryALTinterwordspacing
CTSRD, ``{{CTSRD-CHERI}}/llvm-project,'' Capability Hardware Enhanced RISC
  Instructions, Jul. 2024. [Online]. Available:
  \url{https://github.com/CTSRD-CHERI/llvm-project}
\BIBentrySTDinterwordspacing

\bibitem{Almatary22}
\BIBentryALTinterwordspacing
H.~Almatary \emph{et~al.}, ``{{CompartOS}}: {{CHERI Compartmentalization}} for
  {{Embedded Systems}},'' Jun. 2022. [Online]. Available:
  \url{http://arxiv.org/abs/2206.02852}
\BIBentrySTDinterwordspacing

\bibitem{Conte16}
\BIBentryALTinterwordspacing
M.~Conte, ``Mattconte/tlsf,'' Apr. 2016. [Online]. Available:
  \url{https://github.com/mattconte/tlsf}
\BIBentrySTDinterwordspacing

\bibitem{Watson17}
\BIBentryALTinterwordspacing
R.~N.~M. Watson \emph{et~al.}, ``Capability {{Hardware Enhanced RISC
  Instructions}}: {{CHERI Instruction-Set Architecture}} ({{Version}} 6),''
  Computer Laboratory, University of Cambridge, Technical {{Report}}
  UCAM-CL-TR-907, Apr. 2017. [Online]. Available:
  \url{http://www.cl.cam.ac.uk/techreports/UCAM-CL-TR-907.pdf}
\BIBentrySTDinterwordspacing

\bibitem{Watson19a}
\BIBentryALTinterwordspacing
------, ``Capability {{Hardware Enhanced RISC Instructions}}: {{CHERI
  Instruction-Set Architecture}} ({{Version}} 7),'' Computer Laboratory,
  University of Cambridge, Technical {{Report}} UCAM-CL-TR-927, Jun. 2019.
  [Online]. Available:
  \url{https://www.cl.cam.ac.uk/techreports/UCAM-CL-TR-927.pdf}
\BIBentrySTDinterwordspacing

\bibitem{Ghiti21}
\BIBentryALTinterwordspacing
A.~Ghiti, ``Virtual {{Memory Layout}} on {{RISC-V Linux}},'' Feb. 2021,
  (accessed 2024-07-03). [Online]. Available:
  \url{https://www.kernel.org/doc/html/latest/arch/riscv/vm-layout.html}
\BIBentrySTDinterwordspacing

\bibitem{Spickett23}
\BIBentryALTinterwordspacing
D.~Spickett, ``Top {{Byte Ignore For Fun}} and {{Memory Savings}},'' Feb. 2023,
  (accessed 2024-07-08). [Online]. Available:
  \url{https://www.linaro.org/blog/top-byte-ignore-for-fun-and-memory-savings/}
\BIBentrySTDinterwordspacing

\bibitem{Intel24}
\BIBentryALTinterwordspacing
Intel, \emph{Intel {{Architecture Instruction Set Extensions}} and {{Future
  Features}} - {{Programming Reference}}}, Mar. 2024. [Online]. Available:
  \url{https://cdrdv2-public.intel.com/819680/architecture-instruction-set-extensions-programming-reference.pdf}
\BIBentrySTDinterwordspacing

\bibitem{AMD24}
\BIBentryALTinterwordspacing
AMD, \emph{{{AMD64 Architecture Programmer}}'s {{Manual Volume}} 2: {{System
  Programming}}}, Mar. 2024. [Online]. Available:
  \url{https://www.amd.com/content/dam/amd/en/documents/processor-tech-docs/programmer-references/24593.pdf}
\BIBentrySTDinterwordspacing

\bibitem{Maas24}
\BIBentryALTinterwordspacing
M.~Maas and A.~Zabrocki, ``Working {{Draft}} of the {{RISC-V J Extension
  Specification}},'' Jun. 2024. [Online]. Available:
  \url{https://github.com/riscv/riscv-j-extension}
\BIBentrySTDinterwordspacing

\bibitem{Lowry69}
\BIBentryALTinterwordspacing
E.~S. Lowry and C.~W. Medlock, ``Object code optimization,'' \emph{Commun.
  ACM}, vol.~12, no.~1, pp. 13--22, Jan. 1969. [Online]. Available:
  \url{https://dl.acm.org/doi/10.1145/362835.362838}
\BIBentrySTDinterwordspacing

\bibitem{LLVMteam19}
\BIBentryALTinterwordspacing
{LLVM team}, ``{{LLVM}}: Lib/{{Transforms}}/{{Scalar}}/{{Reg2Mem}}.cpp {{Source
  File}},'' Jan. 2019. [Online]. Available:
  \url{https://llvm.org/doxygen/Reg2Mem_8cpp_source.html}
\BIBentrySTDinterwordspacing

\bibitem{Ruchlejmer24}
\BIBentryALTinterwordspacing
S.~Ruchlejmer, ``Secure {{Rewind}} and {{Discard}} on {{ARM Morello}},'' Jul.
  2024. [Online]. Available: \url{http://arxiv.org/abs/2407.04757}
\BIBentrySTDinterwordspacing

\bibitem{nsajuliet}
{NSA Center for Assured Software}, ``Juliet test suite,''
  \url{https://samate.nist.gov/SARD/test-suites/112}, 2017, accessed:
  2024-04-19.

\bibitem{CTSRD24a}
\BIBentryALTinterwordspacing
CTSRD, ``{{CTSRD-CHERI}}/{{Flute}},'' Capability Hardware Enhanced RISC
  Instructions, Jun. 2024. [Online]. Available:
  \url{https://github.com/CTSRD-CHERI/Flute}
\BIBentrySTDinterwordspacing

\bibitem{Nikhil04}
\BIBentryALTinterwordspacing
R.~Nikhil, ``Bluespec {{System Verilog}}: Efficient, correct {{RTL}} from high
  level specifications,'' in \emph{Proceedings. {{Second ACM}} and {{IEEE
  International Conference}} on {{Formal Methods}} and {{Models}} for
  {{Co-Design}}, 2004. {{MEMOCODE}} '04.}, Jun. 2004, pp. 69--70. [Online].
  Available: \url{https://ieeexplore.ieee.org/document/1459818}
\BIBentrySTDinterwordspacing

\bibitem{Newsome24}
\BIBentryALTinterwordspacing
T.~Newsome \emph{et~al.}, ``Riscv-software-src/riscv-tests,'' RISC-V
  International, Jul. 2024. [Online]. Available:
  \url{https://github.com/riscv-software-src/riscv-tests}
\BIBentrySTDinterwordspacing

\bibitem{MITRE06}
\BIBentryALTinterwordspacing
MITRE, ``{{CWE-457}}: {{Use}} of {{Uninitialized Variable}} (4.14),'' Jul.
  2006, (accessed 2024-07-09). [Online]. Available:
  \url{https://cwe.mitre.org/data/definitions/457.html}
\BIBentrySTDinterwordspacing

\bibitem{Schwartz10}
\BIBentryALTinterwordspacing
E.~J. Schwartz, T.~Avgerinos, and D.~Brumley, ``All {{You Ever Wanted}} to
  {{Know}} about {{Dynamic Taint Analysis}} and {{Forward Symbolic Execution}}
  (but {{Might Have Been Afraid}} to {{Ask}}),'' in \emph{2010 {{IEEE
  Symposium}} on {{Security}} and {{Privacy}}}, May 2010, pp. 317--331.
  [Online]. Available: \url{https://ieeexplore.ieee.org/document/5504796}
\BIBentrySTDinterwordspacing

\bibitem{Podhradsky22}
\BIBentryALTinterwordspacing
M.~Podhradsky, R.~Tadros, and A.~Roach, ``{{GaloisInc}}/{{BESSPIN-GFE}},''
  Galois, Inc., Oct. 2022. [Online]. Available:
  \url{https://github.com/GaloisInc/BESSPIN-GFE}
\BIBentrySTDinterwordspacing

\bibitem{EEMBC24}
\BIBentryALTinterwordspacing
EEMBC, ``Eembc/coremark,'' Embedded Microprocessor Benchmark Consortium, Jul.
  2024. [Online]. Available: \url{https://github.com/eembc/coremark}
\BIBentrySTDinterwordspacing

\bibitem{GCCdevelopercommunity14}
\BIBentryALTinterwordspacing
{GCC developer community}, \emph{Using the {{GNU Compiler Collection For GCC}}
  Version 14.1.0}, May 2014. [Online]. Available:
  \url{https://gcc.gnu.org/onlinedocs/gcc-14.1.0/gcc.pdf}
\BIBentrySTDinterwordspacing

\bibitem{LLVMteam24}
\BIBentryALTinterwordspacing
{LLVM team}, ``Diagnostic flags in {{Clang}},'' Jun. 2024, (accessed
  2024-07-10). [Online]. Available:
  \url{https://releases.llvm.org/18.1.8/tools/clang/docs/DiagnosticsReference.html}
\BIBentrySTDinterwordspacing

\bibitem{Seward05}
\BIBentryALTinterwordspacing
J.~Seward and N.~Nethercote, ``Using {{Valgrind}} to detect undefined value
  errors with bit-precision,'' in \emph{2005 {{USENIX Annual Technical
  Conference}} ({{USENIX ATC}} 05)}.\hskip 1em plus 0.5em minus 0.4em\relax
  Anaheim, CA: USENIX Association, Apr. 2005. [Online]. Available:
  \url{https://www.usenix.org/conference/2005-usenix-annual-technical-conference/using-valgrind-detect-undefined-value-errors-bit}
\BIBentrySTDinterwordspacing

\bibitem{Bruening11}
\BIBentryALTinterwordspacing
D.~Bruening and Q.~Zhao, ``Practical memory checking with {{Dr}}. {{Memory}},''
  in \emph{International {{Symposium}} on {{Code Generation}} and
  {{Optimization}} ({{CGO}} 2011)}, Apr. 2011, pp. 213--223. [Online].
  Available: \url{https://ieeexplore.ieee.org/document/5764689}
\BIBentrySTDinterwordspacing

\bibitem{Stepanov15}
\BIBentryALTinterwordspacing
E.~Stepanov and K.~Serebryany, ``{{MemorySanitizer}}: Fast detector of
  uninitialized memory use in {{C}}++,'' in \emph{Proceedings of the 13th
  {{Annual IEEE}}/{{ACM International Symposium}} on {{Code Generation}} and
  {{Optimization}}}, ser. {{CGO}} '15.\hskip 1em plus 0.5em minus 0.4em\relax
  USA: IEEE Computer Society, Feb. 2015, pp. 46--55. [Online]. Available:
  \url{https://doi.org/10.1145/1250734.1250736}
\BIBentrySTDinterwordspacing

\bibitem{Berger06}
\BIBentryALTinterwordspacing
E.~D. Berger and B.~G. Zorn, ``{{DieHard}}: Probabilistic memory safety for
  unsafe languages,'' \emph{ACM SIGPLAN Notices}, vol.~41, no.~6, pp. 158--168,
  Jun. 2006. [Online]. Available: \url{https://doi.org/10.1145/1133255.1134000}
\BIBentrySTDinterwordspacing

\bibitem{Cao19}
\BIBentryALTinterwordspacing
M.~Cao \emph{et~al.}, ``Different is {{Good}}: {{Detecting}} the {{Use}} of
  {{Uninitialized Variables}} through {{Differential Replay}},'' in
  \emph{Proceedings of the 2019 {{ACM SIGSAC Conference}} on {{Computer}} and
  {{Communications Security}}}, ser. {{CCS}} '19.\hskip 1em plus 0.5em minus
  0.4em\relax New York, NY, USA: Association for Computing Machinery, Nov.
  2019, pp. 1883--1897. [Online]. Available:
  \url{https://doi.org/10.1145/3319535.3345654}
\BIBentrySTDinterwordspacing

\bibitem{Chow05}
J.~Chow \emph{et~al.}, ``Shredding {{Your Garbage}}: {{Reducing Data Lifetime
  Through Secure Deallocation}},'' 2005.

\bibitem{Lu16}
\BIBentryALTinterwordspacing
K.~Lu \emph{et~al.}, ``{{UniSan}}: {{Proactive Kernel Memory Initialization}}
  to {{Eliminate Data Leakages}},'' in \emph{Proceedings of the 2016 {{ACM
  SIGSAC Conference}} on {{Computer}} and {{Communications Security}}}, ser.
  {{CCS}} '16.\hskip 1em plus 0.5em minus 0.4em\relax New York, NY, USA:
  Association for Computing Machinery, Oct. 2016, pp. 920--932. [Online].
  Available: \url{https://doi.org/10.1145/2976749.2978366}
\BIBentrySTDinterwordspacing

\bibitem{Popov18}
\BIBentryALTinterwordspacing
A.~Popov, ``How {{STACKLEAK}} improves {{Linux}} kernel security,'' Nov. 2018,
  (accessed 2024-07-05). [Online]. Available:
  \url{https://a13xp0p0v.github.io/2018/11/04/stackleak.html}
\BIBentrySTDinterwordspacing

\bibitem{Guelton23}
\BIBentryALTinterwordspacing
S.~Guelton, ``Trivial {{Auto Var Init Experiments}},'' Dec. 2023, (accessed
  2024-06-30). [Online]. Available:
  \url{https://serge-sans-paille.github.io/pythran-stories/trivial-auto-var-init-experiments.html}
\BIBentrySTDinterwordspacing

\bibitem{Nilsson23}
\BIBentryALTinterwordspacing
H.~P. Nilsson, ``Bug 111523 -- {{Unexpected}} performance regression with
  -ftrivial-auto-var-init=zero for e.g. systemctl unmask,'' Sep. 2023,
  (accessed 2024-07-07). [Online]. Available:
  \url{https://gcc.gnu.org/bugzilla/show_bug.cgi?id=111523}
\BIBentrySTDinterwordspacing

\bibitem{Yutaka15}
\BIBentryALTinterwordspacing
Y.~Yutaka, ``{{AdLint}},'' Aug. 2015, (accessed 2024-05-05). [Online].
  Available: \url{https://sourceforge.net/projects/adlint/}
\BIBentrySTDinterwordspacing

\bibitem{LLVMteam24b}
\BIBentryALTinterwordspacing
{LLVM team}, ``Clang-{{Check}},'' 2024, (accessed 2024-06-30). [Online].
  Available: \url{https://clang.llvm.org/docs/ClangCheck.html}
\BIBentrySTDinterwordspacing

\bibitem{LLVMTeam24a}
\BIBentryALTinterwordspacing
------, ``Clang-{{Tidy}},'' 2024, (accessed 2024-05-05). [Online]. Available:
  \url{https://clang.llvm.org/extra/clang-tidy/}
\BIBentrySTDinterwordspacing

\bibitem{CodeSecure23}
\BIBentryALTinterwordspacing
CodeSecure, ``{{CodeSonar}},'' May 2023, (accessed 2024-05-05). [Online].
  Available: \url{https://codesecure.com/our-products/codesonar/}
\BIBentrySTDinterwordspacing

\bibitem{Synopsys23}
\BIBentryALTinterwordspacing
Synopsys, ``Coverity {{Scan}},'' Nov. 2023, (accessed 2024-05-05). [Online].
  Available: \url{https://scan.coverity.com/}
\BIBentrySTDinterwordspacing

\bibitem{Marjamaki24}
\BIBentryALTinterwordspacing
D.~Marjam{\"a}ki, ``Cppcheck,'' May 2024, (accessed 2024-05-05). [Online].
  Available: \url{https://www.cppcheck.com/}
\BIBentrySTDinterwordspacing

\bibitem{Wheeler07}
\BIBentryALTinterwordspacing
D.~A. Wheeler, ``Flawfinder,'' Jan. 2007, (accessed 2024-05-05). [Online].
  Available: \url{https://dwheeler.com/flawfinder/}
\BIBentrySTDinterwordspacing

\bibitem{Kirchner15}
\BIBentryALTinterwordspacing
F.~Kirchner \emph{et~al.}, ``Frama-{{C}}: {{A}} software analysis
  perspective,'' \emph{Formal Aspects of Computing}, vol.~27, no.~3, pp.
  573--609, May 2015. [Online]. Available:
  \url{https://doi.org/10.1007/s00165-014-0326-7}
\BIBentrySTDinterwordspacing

\bibitem{Brat14}
G.~Brat \emph{et~al.}, ``{{IKOS}}: {{A Framework}} for {{Static Analysis
  Based}} on {{Abstract Interpretation}},'' in \emph{Software {{Engineering}}
  and {{Formal Methods}}}, D.~Giannakopoulou and G.~Sala{\"u}n, Eds.\hskip 1em
  plus 0.5em minus 0.4em\relax Cham: Springer International Publishing, 2014,
  pp. 271--277.

\bibitem{Facebook16}
\BIBentryALTinterwordspacing
Facebook, ``Infer {{Static Analyzer}},'' May 2016, (accessed 2024-05-05).
  [Online]. Available: \url{https://fbinfer.com/}
\BIBentrySTDinterwordspacing

\bibitem{Evans02}
\BIBentryALTinterwordspacing
D.~Evans, ``Splint,'' Jan. 2002, (accessed 2024-05-05). [Online]. Available:
  \url{https://splint.org/}
\BIBentrySTDinterwordspacing

\bibitem{Rice53}
\BIBentryALTinterwordspacing
H.~G. Rice, ``Classes of recursively enumerable sets and their decision
  problems,'' \emph{Transactions of the American Mathematical Society},
  vol.~74, no.~2, pp. 358--366, 1953. [Online]. Available:
  \url{https://www.ams.org/tran/1953-074-02/S0002-9947-1953-0053041-6/}
\BIBentrySTDinterwordspacing

\bibitem{Bruening04}
\BIBentryALTinterwordspacing
D.~L. Bruening, ``Efficient, {{Transparent}}, and {{Comprehensive Runtime Code
  Manipulation}},'' Ph.D. dissertation, Massachusetts Institute of Technology,
  Sep. 2004. [Online]. Available:
  \url{https://www.burningcutlery.com/derek/phd.html}
\BIBentrySTDinterwordspacing

\bibitem{Nethercote07}
\BIBentryALTinterwordspacing
N.~Nethercote and J.~Seward, ``Valgrind: A framework for heavyweight dynamic
  binary instrumentation,'' in \emph{Proceedings of the 28th {{ACM SIGPLAN
  Conference}} on {{Programming Language Design}} and {{Implementation}}}, ser.
  {{PLDI}} '07.\hskip 1em plus 0.5em minus 0.4em\relax New York, NY, USA:
  Association for Computing Machinery, Jun. 2007, pp. 89--100. [Online].
  Available: \url{https://doi.org/10.1145/1250734.1250746}
\BIBentrySTDinterwordspacing

\bibitem{Cox06}
\BIBentryALTinterwordspacing
B.~Cox \emph{et~al.}, ``N-{{Variant Systems A Secretless Framework}} for
  {{Security}} through {{Diversity}},'' in \emph{15th {{USENIX Security
  Symposium}} ({{USENIX Security}} '06)}, ser. {{USENIX Security}} '06.\hskip
  1em plus 0.5em minus 0.4em\relax Vancouver, B.C. Canada: USENIX Association,
  2006, p.~16. [Online]. Available:
  \url{https://www.usenix.org/conference/15th-usenix-security-symposium/n-variant-systems-secretless-framework-security-through}
\BIBentrySTDinterwordspacing

\bibitem{Salamat09}
\BIBentryALTinterwordspacing
B.~Salamat \emph{et~al.}, ``Orchestra: Intrusion detection using parallel
  execution and monitoring of program variants in user-space,'' in
  \emph{Proceedings of the 4th {{ACM European}} Conference on {{Computer}}
  Systems}, ser. {{EuroSys}} '09.\hskip 1em plus 0.5em minus 0.4em\relax New
  York, NY, USA: Association for Computing Machinery, Apr. 2009, pp. 33--46.
  [Online]. Available: \url{https://doi.org/10.1145/1519065.1519071}
\BIBentrySTDinterwordspacing

\bibitem{Jackson10}
\BIBentryALTinterwordspacing
T.~Jackson, C.~Wimmer, and M.~Franz, ``Multi-variant program execution for
  vulnerability detection and analysis,'' in \emph{Proceedings of the {{Sixth
  Annual Workshop}} on {{Cyber Security}} and {{Information Intelligence
  Research}}}, ser. {{CSIIRW}} '10.\hskip 1em plus 0.5em minus 0.4em\relax New
  York, NY, USA: Association for Computing Machinery, Apr. 2010, pp. 1--4.
  [Online]. Available: \url{https://doi.org/10.1145/1852666.1852708}
\BIBentrySTDinterwordspacing

\bibitem{Koning16}
\BIBentryALTinterwordspacing
K.~Koning, H.~Bos, and C.~Giuffrida, ``Secure and {{Efficient Multi-Variant
  Execution Using Hardware-Assisted Process Virtualization}},'' in \emph{2016
  46th {{Annual IEEE}}/{{IFIP International Conference}} on {{Dependable
  Systems}} and {{Networks}} ({{DSN}})}, Jun. 2016, pp. 431--442. [Online].
  Available: \url{https://ieeexplore.ieee.org/document/7579761}
\BIBentrySTDinterwordspacing

\bibitem{Volckaert16}
S.~Volckaert \emph{et~al.}, ``Secure and efficient application monitoring and
  replication,'' in \emph{Proceedings of the 2016 {{USENIX Conference}} on
  {{Usenix Annual Technical Conference}}}, ser. {{USENIX ATC}} '16.\hskip 1em
  plus 0.5em minus 0.4em\relax USA: USENIX Association, Jun. 2016, pp.
  167--179.

\bibitem{Coppens18}
\BIBentryALTinterwordspacing
B.~Coppens, B.~De~Sutter, and S.~Volckaert, ``Multi-variant execution
  environments,'' in \emph{The {{Continuing Arms Race}}: {{Code-Reuse Attacks}}
  and {{Defenses}}}.\hskip 1em plus 0.5em minus 0.4em\relax {Association for
  Computing Machinery and Morgan \& Claypool}, Mar. 2018, vol.~18, pp.
  211--258. [Online]. Available: \url{https://doi.org/10.1145/3129743.3129752}
\BIBentrySTDinterwordspacing

\bibitem{Osterlund19}
\BIBentryALTinterwordspacing
S.~{\"O}sterlund \emph{et~al.}, ``{{kMVX}}: {{Detecting Kernel Information
  Leaks}} with {{Multi-variant Execution}},'' in \emph{Proceedings of the
  {{Twenty-Fourth International Conference}} on {{Architectural Support}} for
  {{Programming Languages}} and {{Operating Systems}}}, ser. {{ASPLOS}}
  '19.\hskip 1em plus 0.5em minus 0.4em\relax New York, NY, USA: Association
  for Computing Machinery, Apr. 2019, pp. 559--572. [Online]. Available:
  \url{https://doi.org/10.1145/3297858.3304054}
\BIBentrySTDinterwordspacing

\bibitem{OpenSSFcontributors24}
\BIBentryALTinterwordspacing
{OpenSSF contributors}, ``Compiler {{Options Hardening Guide}} for {{C}} and
  {{C}}++,'' Jun. 2024, (accessed 2024-06-30). [Online]. Available:
  \url{https://best.openssf.org/Compiler-Hardening-Guides/Compiler-Options-Hardening-Guide-for-C-and-C++}
\BIBentrySTDinterwordspacing

\bibitem{Skorstengaard19}
\BIBentryALTinterwordspacing
L.~Skorstengaard, D.~Devriese, and L.~Birkedal, ``Reasoning about a {{Machine}}
  with {{Local Capabilities}}: {{Provably Safe Stack}} and {{Return Pointer
  Management}},'' \emph{ACM Trans. Program. Lang. Syst.}, vol.~42, no.~1, pp.
  5:1--5:53, Dec. 2019. [Online]. Available:
  \url{https://doi.org/10.1145/3363519}
\BIBentrySTDinterwordspacing

\bibitem{Serebryany18}
\BIBentryALTinterwordspacing
K.~Serebryany \emph{et~al.}, ``Memory {{Tagging}} and how it improves
  {{C}}/{{C}}++ memory safety,'' Feb. 2018. [Online]. Available:
  \url{http://arxiv.org/abs/1802.09517}
\BIBentrySTDinterwordspacing

\bibitem{Klabnik18}
S.~Klabnik and C.~Nichols, \emph{The {{Rust Programming Language}}}.\hskip 1em
  plus 0.5em minus 0.4em\relax USA: No Starch Press, May 2018.

\bibitem{Silva24}
\BIBentryALTinterwordspacing
T.~Silva, J.~Bispo, and T.~Carvalho, ``Foundations for a {{Rust-Like Borrow
  Checker}} for {{C}},'' in \emph{Proceedings of the 25th {{ACM
  SIGPLAN}}/{{SIGBED International Conference}} on {{Languages}},
  {{Compilers}}, and {{Tools}} for {{Embedded Systems}}}.\hskip 1em plus 0.5em
  minus 0.4em\relax Copenhagen Denmark: ACM, Jun. 2024, pp. 155--165. [Online].
  Available: \url{https://dl.acm.org/doi/10.1145/3652032.3657579}
\BIBentrySTDinterwordspacing

\bibitem{Sutter19}
\BIBentryALTinterwordspacing
H.~Sutter, ``Lifetime safety: {{Preventing}} common dangling,'' Microsoft,
  Technical {{Report}} P1179 R1 -- version 1.1, Nov. 2019. [Online]. Available:
  \url{https://www.open-std.org/jtc1/sc22/wg21/docs/papers/2019/p1179r1.pdf}
\BIBentrySTDinterwordspacing

\bibitem{Lado24}
\BIBentryALTinterwordspacing
Lado, ``Ladroid/{{CppBorrowChecker}},'' Jun. 2024. [Online]. Available:
  \url{https://github.com/ladroid/CppBorrowChecker}
\BIBentrySTDinterwordspacing

\bibitem{Baxter24}
\BIBentryALTinterwordspacing
S.~Baxter and C.~Mazakas, ``Safe {{C}}++,'' Sep. 2024. [Online]. Available:
  \url{https://www.open-std.org/jtc1/sc22/wg21/docs/papers/2024/p3390r0.html}
\BIBentrySTDinterwordspacing

\end{thebibliography}
\onecolumn
\appendices
\section{Technical Supplements}

\subsection{Example of Avoided Data Hazard}
\label{app:datahazard}

% This is the relevant environment
\newenvironment{LastLineToRight}%
  {\setlength{\parindent}{0pt}\setlength{\leftskip}{0pt plus 1fil}\setlength{\rightskip}{0pt plus -1fil}}{\par}

\begin{minipage}{.4\textwidth}
\flushleft
\begin{lstlisting}[style=CStyle, linewidth=.9\columnwidth, title={}, caption={}, label={lst:datahazard}]
/* ... */
%*\dWdth*)csetbounds      ca0, ca0, 4
%*\dCOne*)csetwbrbound    ca0, ca0, zero
%*\dWdth*)li              a1, 10
%*\dCTwo*)csw             a1, 0(ca0)
%*\dCThree*)clw             a0, 0(ca0)
%*\dWdth*)clc             cra, 16(csp)
/* ... */
\end{lstlisting}
  \hspace{5pt}
\end{minipage}% This must go next to `\end{minipage}`
\begin{minipage}{.57\textwidth}
\begin{LastLineToRight}
Listing 2: Example of instruction sequence causing data hazard as described in \Cref{sec:implementation}. At \dCOne, the \texttt{csetwbrbound} instructions sets the \WriteBeforeRead bound for the capability in register \texttt{ca0}, turning it into a \CC{}. At \dCTwo, the capability store word (\texttt{csw}) instructions performs a store operation on \texttt{ca0}, whereupon the operation bound of the capability in \texttt{ca0} is increased at Stage-2 of the pipeline. At the same time, the capability load word (\texttt{clw}) instruction at \dCThree has already 
\end{LastLineToRight}
\vspace{2pt}
\end{minipage}
entered the pipeline and prepares a read via the same capability in \texttt{ca0}. Without the the bypass allowing the updated operation bound to be forwarded from from Stage-2 to Stage-1, this instruction would fail due to the operation bound check on the out-of-date bound. With the bypass, this sequence of instructions is valid and does not incur additional latency.
\addtocounter{lstlisting}{1}

\subsection{Changes to Capability Encoding}
\label{app:decoding}

\begin{figure}[h!]

\resizebox{\textwidth}{!}{\usebox{\opboundscompressed}}

\vspace{-10pt}
\begin{center}
  \scriptsize
  \begin{tabular}{p{0.1\linewidth}p{0.15\linewidth}p{0.25\linewidth}p{0.2\linewidth}p{0.15\linewidth}}
    \CCFlags: flag & \CCPermissions: permissions & \CCOpPerms: \CP control bits &\CCOType: object type & \CCAddress: pointer address
  \end{tabular}
\end{center}

\vspace{2pt}
\begin{center}
  \scriptsize
  %\begin{tabular}{p{0.4\linewidth}|p{0.4\linewidth}}
  \begin{tabular}{c c| c c}
    \multicolumn{1}{l}{If $\CCInternalExponent{} = 0$:}
    & &
    \multicolumn{1}{l}{If $\CCInternalExponent{} = 1 \textrm{ \newtext{and} } \newmath{\CCExponent = 1,2:}$} &\\

    & 

    {$\begin{aligned}
        \CCExponent &= 0 \\
        \CCBottomBitsTop &= \CCExpHighPart{} \\
        \CCBottomBitsBase &= \CCExpLowPart{} \\
        \newmath \CCBottomBitsOp & \newmath = \newmath \CCExpZeroExtOpHighPart \\
        \LCarry &= \begin{cases}%
            \begin{aligned}
                1,\quad&\textrm{if } \CCExpZeroTop < \CCExpZeroBaseBitsToCompare \\
                0,\quad&\textrm{otherwise}
            \end{aligned}
        \end{cases} \\
        \LMsb &= 0 \\
    \end{aligned}$}

    & &
    
    {$\begin{aligned}
       \CCExponent &= \left\{\CCExpHighPart, \CCExpLowPart\right\} \\
        \CCBottomBitsTop &= 0 \\
        \CCBottomBitsBase &= 0 \\
        \newmath \CCBottomBitsOp & \newmath = \newmath{\CCExtOp\left[E+2:E\right]} \\
        \LCarry &= \begin{cases}%
          \begin{aligned}
            1,\quad&\textrm{if } \CCExpNonZeroTop < \CCExpNonZeroBase \\
            0,\quad&\textrm{otherwise}
          \end{aligned}
        \end{cases} \\
        \LMsb &= 1 \\
    \end{aligned}$}

    \\
  \end{tabular}
\end{center}

\vspace{8pt}
Reconstituting the top two bits of $T$:

\vspace{-18pt}
\begin{center}
\scriptsize
\[
    \begin{split}
         \CCUpperBitsTop & = \CCUpperBitsBase + \LCarry + \LMsb \\
    \end{split}
\]
\end{center}

Decoding the bounds:

\begin{center}
  \scriptsize
  \begin{tabular}{r|c|c|c|}
      address, $a =$ & $ a_{top} = a\left[\newmath{47} : \CCExponent + 14\right]$ & $a_{mid} = a\left[\CCExponent + 13: \CCExponent\right] $ & $ a_{low} = a\left[\CCExponent -1: 0 \right] $ \\
      top,      $t =$ & $ a_{top} + c_t $ & $ \CCAllBitsTop    $ & $0 '\CCExponent $ \\
      bottom,   $b =$ & $ a_{top} + c_b $ & $ \CCAllBitsBase $ & $0 '\CCExponent $ \\
      \newtext{operation $o =$} & $ \newmath{a_{top} + c_o} $ & $ \newmath{\CCAllBitsOp} $ & 

    {$\begin{aligned}
      \newmath{\textrm{If }\CCInternalExponent} & \newmath = \newmath 0: \\
      & \newmath{O'E} \\
      \newmath{\textrm{If }\CCInternalExponent} & \newmath = \newmath 1 \textrm{ \newtext{and} } \newmath{\CCExponent = 1,2:} \\
      & \newmath{\CCExtOp\left[E-1:0\right] } \\
    \end{aligned}$}
      \\
  \end{tabular}
\end{center}

%\vspace{12pt}

To calculate the corrections $c_t$, $c_b $ \newtext{and $c_o$}:

\vspace{-18pt}
\begin{center}
\scriptsize
    {\begin{align}
      A_3 &= \CCAddress\left[E+13:E+11\right] \\
        B_3 &= \CCBase\left[13:11\right] \\
        T_3 &= \CCTop\left[13:11\right]\\
        \newmath{O_3} &= \newmath{\CCOp\left[13:11\right] } \\
        R &= B_3 - 1
    \end{align}}
\end{center}

%\vspace{12pt}
\begin{center}
\scriptsize
\begin{tabular}{ccr}\toprule
  $A_3 < R$ & $ T_3 < R $ & $ c_t $ \\
  \midrule
  false & false & $0$  \\
  false & true  & $+1$ \\
  true  & false & $-1$ \\
  true  & true  & $0$  \\
  \bottomrule
\end{tabular}
\quad
\begin{tabular}{ccr}\toprule
  $A_3 < R$ & $ B_3 < R $ & $ c_b $ \\
  \midrule
  false & false & $0$  \\
  false & true  & $+1$ \\
  true  & false & $-1$ \\
  true  & true  & $0$  \\
  \bottomrule
\end{tabular}
\quad
{\color{red}%
\begin{tabular}{ccr}\toprule
  $A_3 < R$ & $ O_3 < R $ & $ c_o $ \\
  \midrule
  false & false & $0$  \\
  false & true  & $+1$ \\
  true  & false & $-1$ \\
  true  & true  & $0$  \\
  \bottomrule
\end{tabular}}
\end{center}

\caption{Compressed 128-bit capability format and decoding (adapted from~\cite{Watson23} with \newtext{additions and changes marked in red}).}\label{fig:format}

\end{figure}
\newpage

\subsection{Examples from the Juliet Test Suite}
\label{app:juliet}

\begin{lstlisting}[float=h!, style=CStyle, label={lst:juliet-cwe457-63-good},
caption={A ``good'' example from the Juliet Test Suite~\cite{nsajuliet} where \moncheri detects an uninitialized variable use. A reference to an uninitialized \texttt{volatile double data} \dCOne is passed in a function call \dCTwo and dereferenced and copied \dCThree in the callee where \moncheri detects an uninitialized load. The memory content of the copy is initialized before use in \dCFour. The example is from \texttt{C/testcases/CWE457\_Use\_of\_Uninitialized\_Variable/s01/CWE457\_Use\_of\_Uninitialized\_Variable\_\_double \_63a.c} in the official test suite release.}]
void CWE457_Use_of_Uninitialized_Variable__double_64b_goodB2GSink(void * dataVoidPtr)
{
 %*\dWdth*) /* cast void pointer to a pointer of the appropriate type */
 %*\dWdth*) double * dataPtr = (double *)dataVoidPtr;
 %*\dWdth*) /* dereference dataPtr into data */
 %*\dCThree*) volatile double data = (*dataPtr);
 %*\dWdth*) /* FIX: Ensure data is initialized before use */
 %*\dCFour*) data = 5.0;
 %*\dWdth*) printDoubleLine(data);
}

static void goodB2G()
{
 %*\dCOne*) volatile double data;
 %*\dWdth*) /* POTENTIAL FLAW: Don't initialize data */
 %*\dWdth*) ; /* empty statement needed for some flow variants */
 %*\dCTwo*) CWE457_Use_of_Uninitialized_Variable__double_63b_goodB2GSink(&data);
}

void CWE457_Use_of_Uninitialized_Variable__double_63_good()
{
 %*\dWdth*) goodG2B();
 %*\dWdth*) goodB2G();
}
\end{lstlisting}

  \begin{lstlisting}[style=CStyle, label={lst:64bsink}, caption={An example code from the Juliet Test Suite, demonstrating the use of partially initialized arrays. The function \texttt{CWE457\_Use\_of\_Uninitialized\_Variable\_\_int\_array\_alloca\_partial\_init\_64\_bad} allocates memory for an integer array using \texttt{ALLOCA} \dOne and partially initializes it \dTwo. The partially initialized array is then passed to \texttt{CWE457\_Use\_of\_Uninitialized\_Variable\_\_int\_array\_alloca\_partial\_init\_64b\_badSink} \dThree, and that function is reading the whole extent of the \texttt{data} array. That uninitialized memory access detected by \moncheri{}}.]
__attribute__((writebeforeread, noinline))
void CWE457_Use_of_Uninitialized_Variable__int_array_alloca_partial_init_64_bad()
{
 %*\dWdth*) volatile int *data;
 %*\dOne*) data = (int *)ALLOCA(10*sizeof(int));
 %*\dWdth*) /* POTENTIAL FLAW: Partially initialize data */
 %*\dWdth*) {
 %*\dWdth*)     int i;
 %*\dTwo*)     for(i=0; i<(10/2); i++)
 %*\dWdth*)     {
 %*\dWdth*)         data[i] = i;
 %*\dWdth*)     }
 %*\dWdth*) }
 %*\dThree*) CWE457_Use_of_Uninitialized_Variable__int_array_alloca_partial_init_64b_badSink(&data);
}
\end{lstlisting}

\begin{lstlisting}[style=CStyle, label={lst:64bsinkir}, caption={LLVM IR of \Cref{lst:64bsink} after \CP the store linearization. \dCOne allocates memory on the stack for a pointer. \dCTwo allocates memory for an array of 40 bytes. \dCThree stores the capability \texttt{\%4} at the memory location pointed to by \texttt{\%1}. After \dCFour, the operation bound is updated at the hardware level, so the capability stored in memory becomes invalid. Our instrumentation adds another store to update the capability in memory \dCFive. }]
define dso_local void @CWE457_Use_of_Uninitialized_Variable__int_array_alloca_partial_init_64_bad() local_unnamed_addr addrspace(200) #0 !dbg !19 {
entry:
 %*\dCOne*) %data = alloca ptr addrspace(200), align 16, addrspace(200), !clang.decl.ptr !28
 %*\dWdth*) %0 = call ptr addrspace(200) @llvm.cheri.bounded.stack.cap.i64(ptr addrspace(200) %data, i64 16)
 %*\dWdth*) %1 = call ptr addrspace(200) @llvm.cheri.cap.op.bounds.set.i64(ptr addrspace(200) %0, i64 0)
 %*\dCTwo*) %2 = alloca [40 x i8], align 16, addrspace(200), !dbg !30
 %*\dWdth*) %3 = call ptr addrspace(200) @llvm.cheri.bounded.stack.cap.i64(ptr addrspace(200) %2, i64 40), !dbg !30
 %*\dWdth*) %4 = call ptr addrspace(200) @llvm.cheri.cap.op.bounds.set.i64(ptr addrspace(200) %3, i64 0), !dbg !30
 %*\dCThree*) store ptr addrspace(200) %4, ptr addrspace(200) %1, align 16, !dbg !32, !tbaa !33
 %*\dWdth*) ...
 %*\dCFour*) store volatile i32 0, ptr addrspace(200) %4, align 16, !dbg !38, !tbaa !42
 %*\dCFive*) store volatile ptr addrspace(200) %4, ptr addrspace(200) %1, align 16, !dbg !37
 %*\dWdth*) ...
}
\end{lstlisting}

\ifnotarxiv
\input{sections/meta_review.tex}
\fi

\end{document}